\documentclass[a4paper,11pt]{article}
\usepackage{jinstpub} 
\usepackage{lineno}

\title{\boldmath GAMPix: a novel fine-grained, low-noise and ultra-low power pixelated charge readout for TPCs}

\author[a, b, 1]{Tom Shutt\note{Corresponding author.}}
\author[a, b]{Bahrudin Trbalic}
\author[a]{Aldo Pena-Perez}
\author[a]{Steffen Luitz}
\author[a]{Mark Convery}
\author[a]{Angelo Dragone}
\author[a]{Lorenzo Rota}
\author[a]{Dietrich R. Freytag}
\author[a]{Dionisio Doering}
\author[c, d]{Filippo Mele}
\author[a]{Miriam Moore}
\author[a]{Hiro Tanaka}
\author[a]{Yun-Tse Tsai}

\affiliation[a]{SLAC National Accelerator Laboratory,\\ Menlo Park, CA 94025, USA}
\affiliation[b]{Kavli Institute for Particle Astrophysics and Cosmology, Stanford University, Stanford, CA 94305-4085 USA}
\affiliation[c]{Politecnico di Milano, Via Anzani, Como, Italy}
\affiliation[d]{INFN-Milano, Via Celoria, Milan, Italy}

\emailAdd{tshutt@slac.stanford.edu}

\abstract{We report on the development of a novel pixel charge readout system, Grid Activated Multi-scale pixel readout (GAMPix), which is under development for use in the GammaTPC gamma ray instrument concept.  GammaTPC is being developed to optimize the use of liquid argon time projection chamber technology for gamma ray astrophysics, for which a fine grained low power charge readout is essential.  GAMPix uses a new architecture with coarse and fine scale instrumented electrodes to solve the twin problems of loss of measured charge after diffusion, and high readout power.  Fundamentally, it enables low noise and ultra low power charge readout at the spatial scale limited by diffusion in a time projection chamber, and has other possibly applications, including 
future DUNE modules.}


\begin{document}
\maketitle
\flushbottom

\section{Introduction}

\subsection{Liquid noble TPCs}

We report on GAMPix, a new fine-grained charge readout architecture we have been developing for the GammaTPC gamma ray instrument concept \cite{gammatpc24}, which uses liquid argon (LAr) time projection chamber (TPC) technology.  GAMPix has other applications in liquid noble (and potentially gaseous) TPCs, including potentially substantially improving low energy readout in the DUNE neutrino experiment~\cite{dunefd2_2023}. Liquid noble TPCs have had a major impact in dark matter searches where liquid Xe (LXe) TPCs have had the leading sensitivity to WIMP dark matter for more than a decade (see, e.g. \cite{LZ2023, xenonnt2023, pandax2023}), and in neutrinos where LAr TPCs are the basis for DUNE and LXe TPCs are a leading technology for neutrinoless $\beta\beta$ searches \cite{nexo_2018}.       

\begin{figure}
    \centering
    \includegraphics[width=1\textwidth]{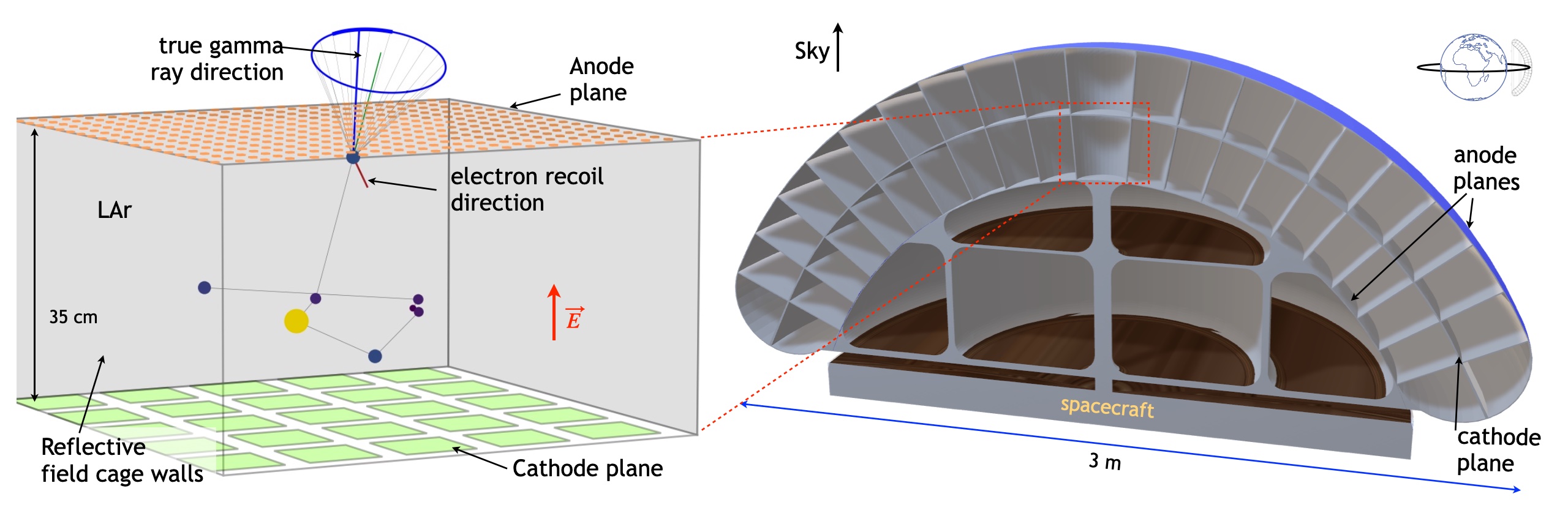}
    \caption{(left) Typical $\gamma$-ray event with multiple scatters in a schematic LAr TPC cell, with the reconstructed event cone and arc shown. (right) 
    Cross section schematic of a large version of the GammaTPC concept.  A 35\,cm thick layer of LAr, 10\,m$^2$ in area and 4 metric tons mass is contained in a th in-walled carbon fiber shell, segmented into two layers of $\sim\,20\,cm$ TPC cells. The geometry is a section of a sphere which provides uniform response over the full sky in the orbit sketched.}
    \label{fig:event_detector}
\end{figure}

In Fig.~\ref{fig:event_detector}~\emph{(left)} we show a 1.0 MeV gamma ray that has undergone seven Compton scatters in LAr, followed by photoabsorption, with the LAr instrumented as a TPC. The simulation was done with the Geant4-based MEGAlib package \cite{MEGAlib}. The cathode plane of the TPC serves as an electrode biased at high negative voltage compared to the anode plane, to establish a 0.5\,kV/cm electric drift field.  Both planes are populated with readout; the cathode with silicon photon multiplier (SiPM) light sensors, and the anode plane with charge readout that is the focus of this article.  The walls have field-grading electrodes to ensure a uniform drift field, and in this case are coated with waveshifter and lined with a polymer film~\cite{vikuiti} that is highly reflective for the wave-shifted light.  

Each of the interactions of gamma ray creates a high energy electron track (10s-100s of keV), which in turn generate scintillation light and free charge (electrons + ions).  The combined light from all scatters is measured as a prompt ($\delta T\,\sim$ 10 ns) signal, and the time difference between this and the arrival of the charge signals from the electron tracks measures the depths of those tracks, and gives rise to the name ``time projection chamber''.  The $x$--$y$ location of the tracks is measured directly by the pixel electrodes in the anode plane. The overall GammaTPC instrument concept is shown in Fig.~\ref{fig:event_detector}~(\emph{right}), where the active volume is segmented into many TPC cells to limit event pile up given the relatively slow drift speed of electrons ($\sim$\,1\,mm/$\mu$s) combined with the high rate of particles in space.

There are several advantages of a TPC, starting with the fact that 3D imaging is achieved using only 2D sensor arrays on the surfaces, which for full pixelated $x$--$y$ readout, reduces $N^3$ channels to $N^2$. The very significant cost and power savings from this allows us in this application to push $N$ to the scale of 10$^3-10^4$.  It also provides a uniform detection medium with dead material only at the periphery. This is useful for different reasons in different contexts.  In large LXe TPCs for dark matter and $\beta\beta$ decay searches it allows for a very low background central region which is self-shielded against external penetrating backgrounds of gammas and neutrons.  It also allows uniform imaging over a large area, as in the extended GeV events which are uniformly measured in fine detail in the DUNE far detector \cite{dunefd2_2023}. For gamma rays it allows uniform measurement of all scatters without loss to dead material (though the GammaTPC application requires segmentation which compromises this benefit).  

The light signal is important for measuring the event energy, and also establishes the event start time (apart from applications such as a pulsed beam which independently sets the start time). The light can be measured with arrays of PMTs, waveshifting bars, or arrays of SiPMs, but discussion of these topics is beyond the scope of this paper.  For completeness we note that measuring the very low energy signals in a dark matter search requires amplifying of the charge signal, to date done with two phase detectors and electroluminescence in the gas phase \cite{Bolozdynya1999}. However most TPCs measure charge directly on electrodes, and this is what we consider here. The principle limitation of TPCs is the relatively slow drift of charge (roughly $1\,mm/\mu s$ in both LXe and LAr, depending on the drift field) prevents their use in high rate applications.  Also, diffusion of the drifting electrons sets a fundamental limit to the spatial resolution for imaging of complex structures. 

\subsection{Requirements GammaTPC charge readout}\label{requirements}

The GAMPix architecture arose from the demanding requirements of the Compton telescope technique which reconstructs the incoming direction of gamma rays for events like that shown in Fig.~\ref{fig:event_detector}. In the figure the photon path is sketched, but in reality must first be deduced as follows.  With the interaction locations $\vec{r_i}$ and deposited energies $E_i$ of all the numerous interactions ($\left< N\right> \sim$ 5, depending on the gamma ray energy), a set of kinematic tests compare the geometric angles formed in triplets of interactions to the angle derived from the energetics of those interactions.  This is done for the combinatorial set of all possible interaction sequences~\cite{Aprile1993, Boggs_2000}, along with other information such as the distribution of distances between scatters and weighting of the scatters to a surface of the detector~\cite{Zoglauer2006} to find the most likely reconstructed sequence. Then the first two scatters and the energetically determined angle of the first scatter, a cone of possible incident gamma ray directions in determined.  With multiple photons from the same source, an image of the source is formed by the overlap of cones from many events. If the initial direction of the first electron recoil can be measured, the cone is reduced to an arc (this would be a point in the implausible limit of perfect direction measurement), sharply improving the efficiency of forming a source image from events.  

\begin{wrapfigure}{l}{0.5\textwidth}
    \centering
    \includegraphics[width=00.48\textwidth]{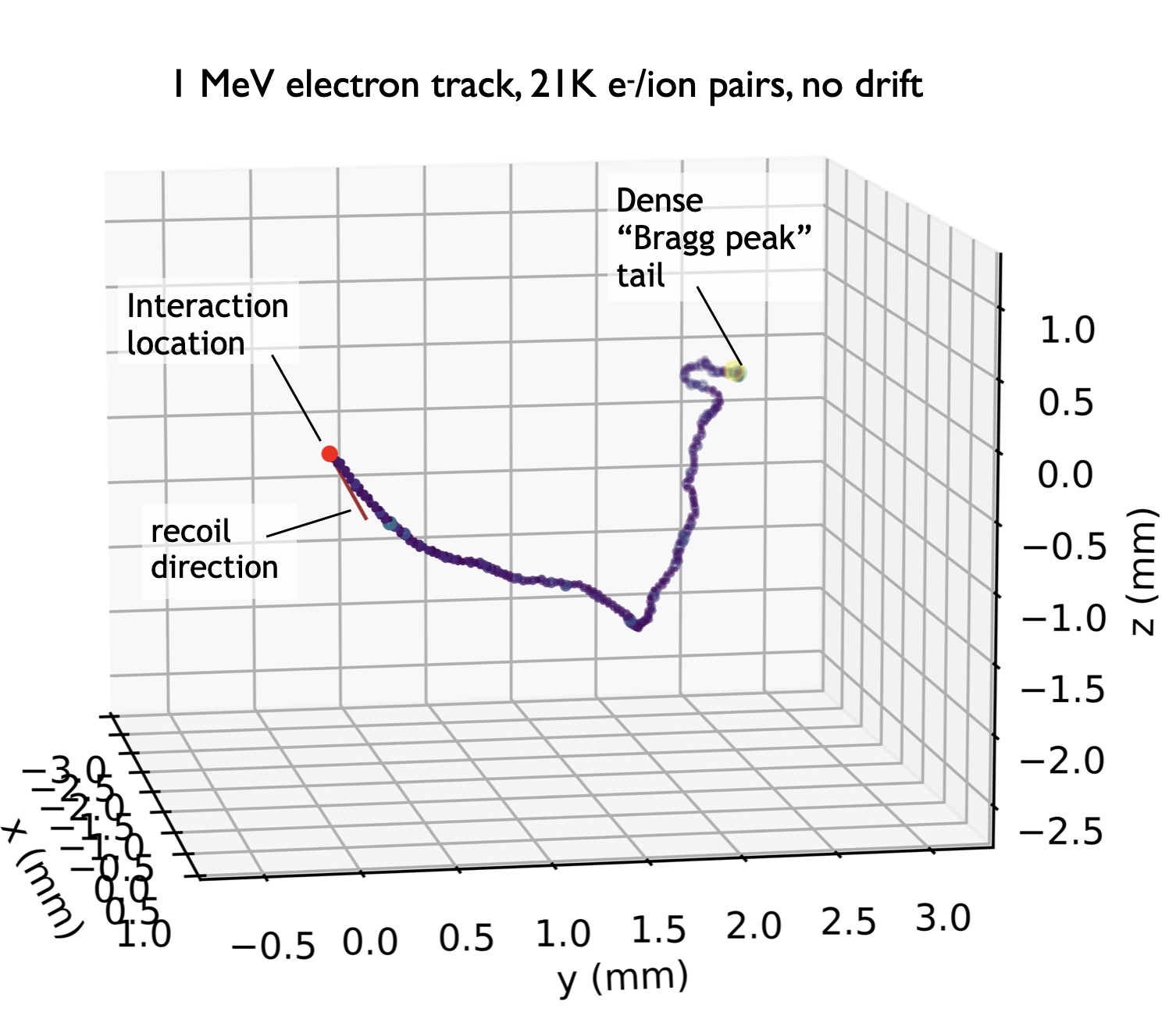}
    \caption{1.0 MeV electron recoil track, prior to drift, voxelized in 10 $\mu$m bins for display. Simulated with PENELOPE \cite{penelope} with the physics tracked to below 500 eV.}
    \label{fig:ElectronTrack}
\end{wrapfigure}

The error on the measurements of the $\left\{E_i \right\}$  and $\left\{ \vec{r_i} \right\}$ are thus crucial, as they alone determine the width of the point-spread function (psf) of the resulting image which for nearly all astrophysical measurements is a direct driver of sensitivity, often as the square of the psf.  In the same way they are crucial to the efficiency of the kinematic sequence reconstruction, with all mis-reconstructed events not only not contributing to forming the image, but by contributing an incorrect circle or arc, become potential background for that image. To be competitive with other technologies~\cite{De_Angelis_2017, Fleischhack_2021, McEnery:2019tcm, cosi_2019}, a spatial resolution at or below 1\,mm is needed.  To further measure the initial electron direction, arguably the ``holy grail'' of the Compton telescope technique, even finer resolution is desirable.  Moreover, this spatial resolution is less than the size of the electron recoil tracks from typical scatters from MeV gammas, with an example track shown in Fig. \ref{fig:ElectronTrack}. Thus such a track must be imaged on a fine enough scale so that the location of the head, or $\vec{r_i}$, can be measured, which also allows the initial direct direction to be measured.  The head and tail are distinguishable, at least in principle, because the tail has higher density due to the Bragg peak from large $dE/dx$ at low energy \cite{pdg2022}.  This level of imaging is also required for higher energy gamma rays that interact via pair production, and for which the direction is determined by high quality imaging of the initial portion of the $e^+e^-$ tracks.

Finally, the need for good energy resolution imposes a requirement that the effective rms charge readout noise be no more than 75\,$e^-$ (noise in $e^-$ is also called ENC or equivalent noise charge), which in energy is equivalent to roughly 3.5\,keV.  All of this must be accomplished in space, where the limitations of cryogenics give a very demanding power budget to the charge readout system of a few W/m$^2$~\cite{gammatpc24}.  

\subsection{Charge readout}\label{charge_readout}

To place our new charge architecture in context, we begin with a brief review of the well known foundations of charge readout (see, for example, \cite{Radeka} and references therein).
A charge readout consists of a set of sense electrodes connected to charge sensitive amplifiers (CSA).  The sense electrodes are often a subset of the electrodes used to generate the electric field which induces charge drift, and the charge sensitive amplifier holds the electrodes at a fixed potential while sensing the movement of nearby charge~\cite{Ramo_39, Shockley1938}. The fine-grained charge readout in a TPC generally uses a set of closely spaced sets of crossed wires or strips, or a 2D array of pixels.  

The noise in the CSA amplifier depends on several factors, including the noise of the CSA's front end FET, and in general scales linearly with the total capacitance to ground (or to fixed voltage sensors) at the input of the CSA.  That capacitance is the sum of the sensor capacitance, the input capacitance of the front end FET, and a parasitic cabling capacitance, or $C_{in} = C_{sense} + C_{FET} + C_p$.  The noise in the FET decreases with increasing FET input capacitance and decreasing temperature $T$. For an amplifier with a discrete Si JFET, with the FET matched to the input sensor capacitance and with negligible $C_p$, the noise scales as $\sqrt{C_{sense}T}$ \cite{Radeka}.  Thus small sensors at cryogenic temperatures have low noise. 

To achieve the lowest noise possible it is essential to minimize any parasitic capacitance, $C_p$, which means locating the CSAs as close as possible to the sensors to minimize the capacitance of the cabling between sensors and amplifiers. This requires the CSA to be housed in the low temperature environment and be capable of running at that temperature. With very large channel counts, it also preferable to use readily scalable CMOS technology. Recently, several CMOS ASICs designed to work at LAr temperatures have been developed for DUNE \cite{Adams_2020}, \cite{Dwyer_2018}, \cite{CryoAsic1}, at LXe temperatures for nEXO \cite{nexo_2018, CryoAsic2, CryoAsic3}, and at or below 4 K for quantum computing and quantum instrumentation (see, e.g., \cite{Absar2024}). The input MOSFETs for CSAs built with CMOS have elevated $1/f$ noise compared to Si JFETs, but nonetheless can achieve low noise levels~\cite{OConnor2000}.  Crucially, low noise requires high power, for many CMOS families, scaling inversely as the square root of the power in the front end FET~\cite{OConnor2000}. Thus there is a fundamental power challenge for applications with many channels of low noise readout, as a factor of 10 noise improvement requires 100 times more power.

\begin{wrapfigure}{r}{0.38\textwidth}
    \centering
    \includegraphics[width=0.35\textwidth]{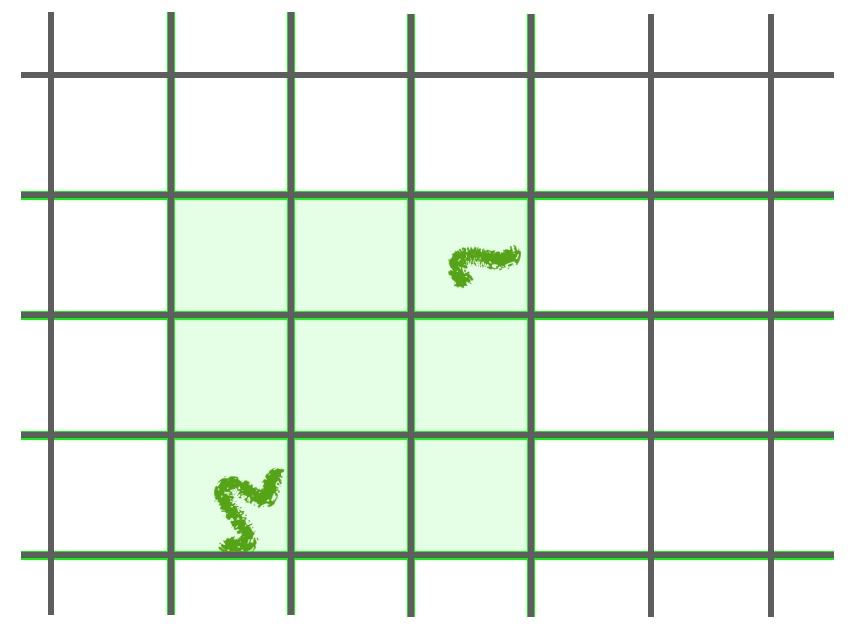}
    \caption{Illustration of ``ghost'' images for two tracks measured with an $x$--$y$ array of wires or strips. The tracks are confined to 2 ``boxes'' between wires, but the signal on 4 sets of $x$ and 4 sets of $y$ wires only constrains the signals to a region of $3 \times 3=9$ boxes.}
    \label{fig:Ghosts}
\end{wrapfigure}

There are a variety of configurations of the readout electrodes, but for full 3D readout a common feature is that they consist of crossed $x$--$y$ or $u$--$v$--$w$ arrays of wires or strips, or a 2D array of pixels.  Wires provide full $x$--$y$ imaging for a single point-like charge distribution, but as shown in section~\ref{fig:Ghosts}, run into the problem of "ghost" images for more complex charge distributions. DUNE readout uses three $u$--$v$--$w$ wire planes, where the third direction substantially reduces but does not fully eliminate this ghosting.  Only pixels provide true 3D imaging without ghosting, and are particularly helpful for the GammaTPC application.  However pixels come at a severe cost in channel count: for sensors at pitch $s$ spanning a detector length detector $L$, only $2N$ sensor wires or strips are needed where $N = L/s$, compared to $N^2$ pixels. Thus pixels require a factor $N/2$ times more readout channels, greatly increasing power and also cost.  For both GammaTPC and DUNE, $N$ is on the scale of thousands. 

\subsection{Two challenges for fine-scale low noise charge readout:}\label{two_challenges}

Even disregarding the cost and complexity associated with a large number of channels, achieving fine-scale, low-noise readout in a liquid noble TPC faces two fundamental challenges, both effectively addressed by the GAMPix architecture. The first is the power at high channel count, particularly for low noise readout which requires higher power per channel. That power is typically 0.1\,-\,10\,mW/ch in the front end FET alone, depending on the input capacitance. The extent to which this power requirement poses a problem is contingent on the specific application. In the GammaTPC example, where desired spatial resolution and a sensor pitch of 0.5\,mm are needed, pixels operating at 0.25\,mW/ch result in a power density of 1\,kW/m$^2$. This power density is approximately three orders of magnitude higher than the power budget allowed by basic considerations of achievable cryogenic cooling in a spacecraft \cite{gammatpc24}.

\begin{figure}[htbp]
  \centering
  \includegraphics[width=0.48\textwidth]{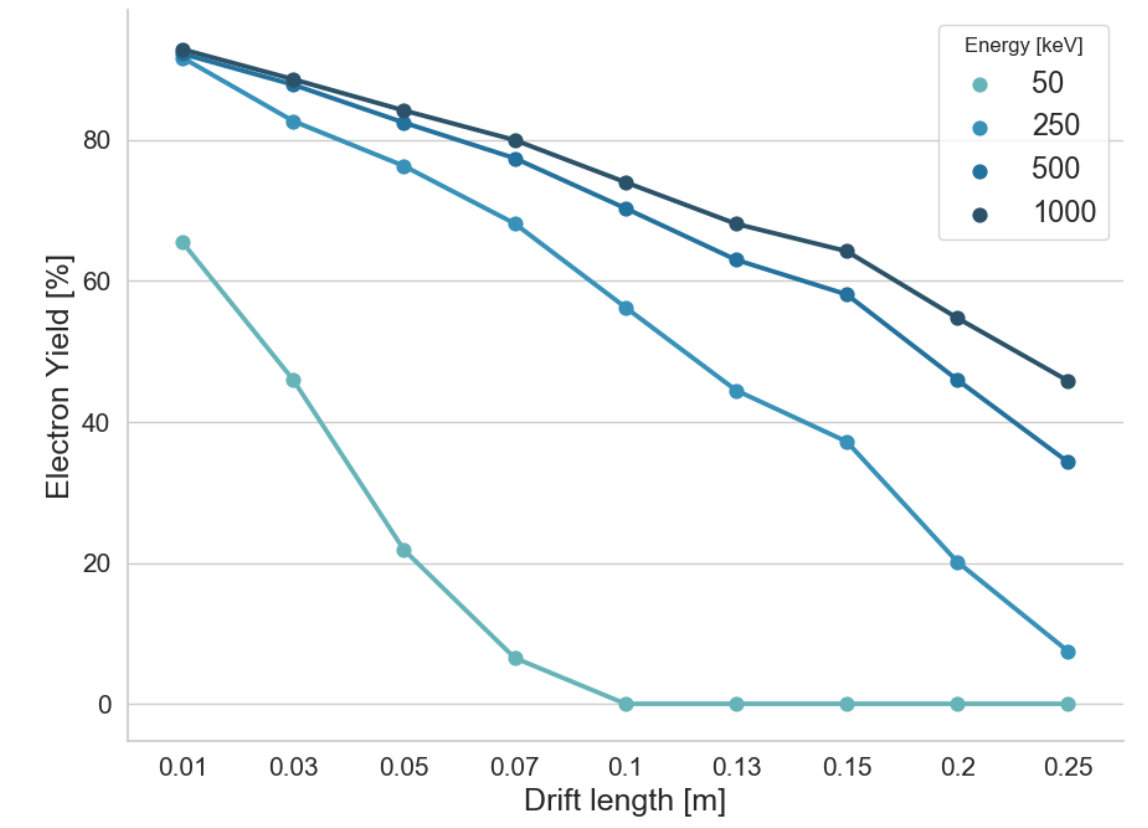}
  \includegraphics[width=0.48\textwidth]{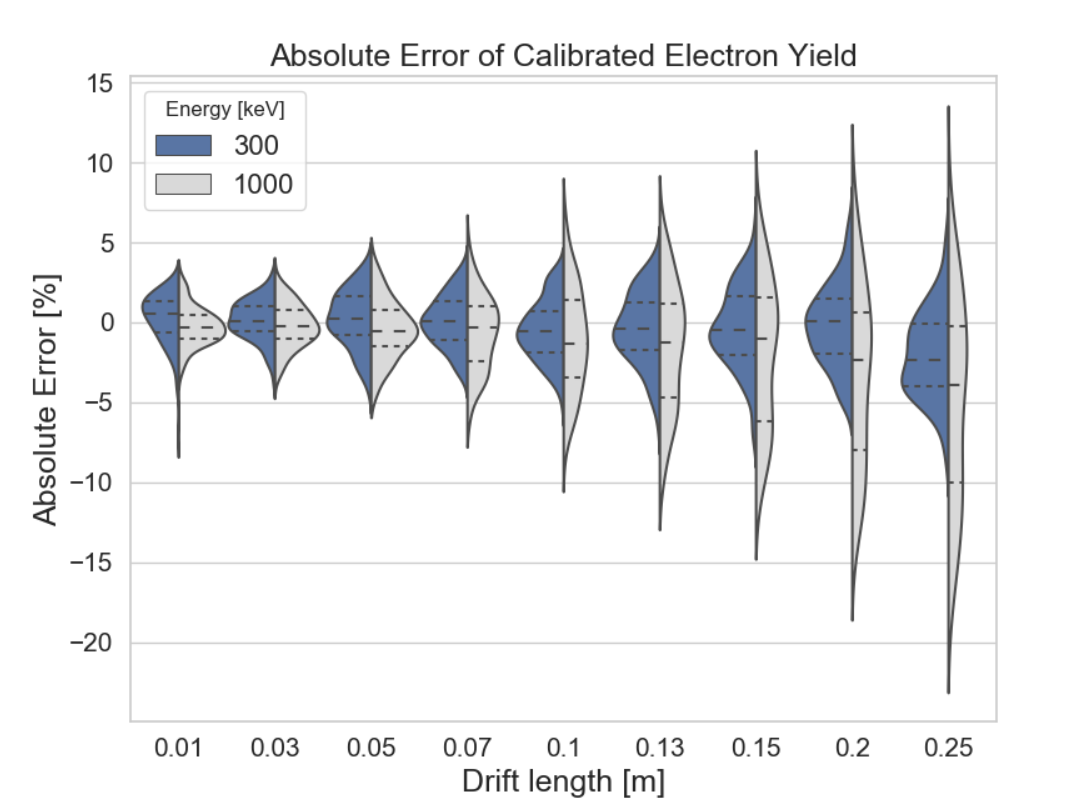}
    \caption{\emph{(left)} Simulation of the mean measured electron yield for electron tracks of the energies listed, measured with 500\,$mu$m  pitch pixels, and 80\,$e^-$ threshold (nominally, a 15\,$e^-$ ENC). \emph{(right)} Resolution on the estimated charge using an ML method to account for the loss due to diffusion, and variations in that loss due to the underlying track shapes. The ML method is trained on a variety of track shapes, energies and drift lengths on a two layer Neural Network. }
    \label{fig:LossDiffusion}
\end{figure}

The second challenge is more subtle but arguably more fundamental, and applies whenever the required spatial resolution scale and hence sensor pitch $s$ is comparable to the scale of diffusion. Diffusion sets an irreducible limit to spatial information in a TPC, and so this condition applies whenever a TPC charge readout is designed to achieve the optimal possible spatial resolution.  The problem in these cases is that charge at the periphery of the the diffused charged distribution will fall below threshold in sensors unless the noise is below one\,$e^{-1}$, which is generally not achievable. This problem is present for wire or strip readout, and is more severe for pixels where the effect occurs in 2 dimensions instead of 1.  We show a simulation of this in Fig.~ \ref{fig:LossDiffusion}~(\emph{left}) for the GammaTPC 500\,$\mu$m pitch pixel readout. At 25\,cm drift, diffusion has $\sigma_D$\,=\,650\,$\mu$m, so this effect does not require the diffusion to be much larger than the pitch for the effect to be profound, especially at low energies.

Because these trends are relatively smooth, this effect can be calibrated and corrected for, but such a correction is incomplete in two important ways illustrated in Fig.~\ref{fig:LossDiffusion}.  First, at sufficiently low energy signals are fully lost to threshold, which is effectively fatal for the GammaTPC application given its threshold requirements.  Second, the amount of missing charge depends on variations in the track shape, with point like tracks losing less charge than extended line-like tracks.  Since the pixel data images the track shapes this can be accounted for, but the process is not perfect.  In the right panel we show the residual variations in the estimated amount of charge from the data in the left panel, using a ML method that uses the pixel-measured track shape information.  The residual variations are much larger than expected from $\sqrt{N}$ statistics, which for GammaTPC would substantially worsen event reconstruction, pointing, and energy resolution.

\section{GAMPix}

\subsection{The GAMPix Architecture}\label{gampix_overview}

\begin{figure}[ht]
    \centering
    \includegraphics[width=0.85\textwidth]{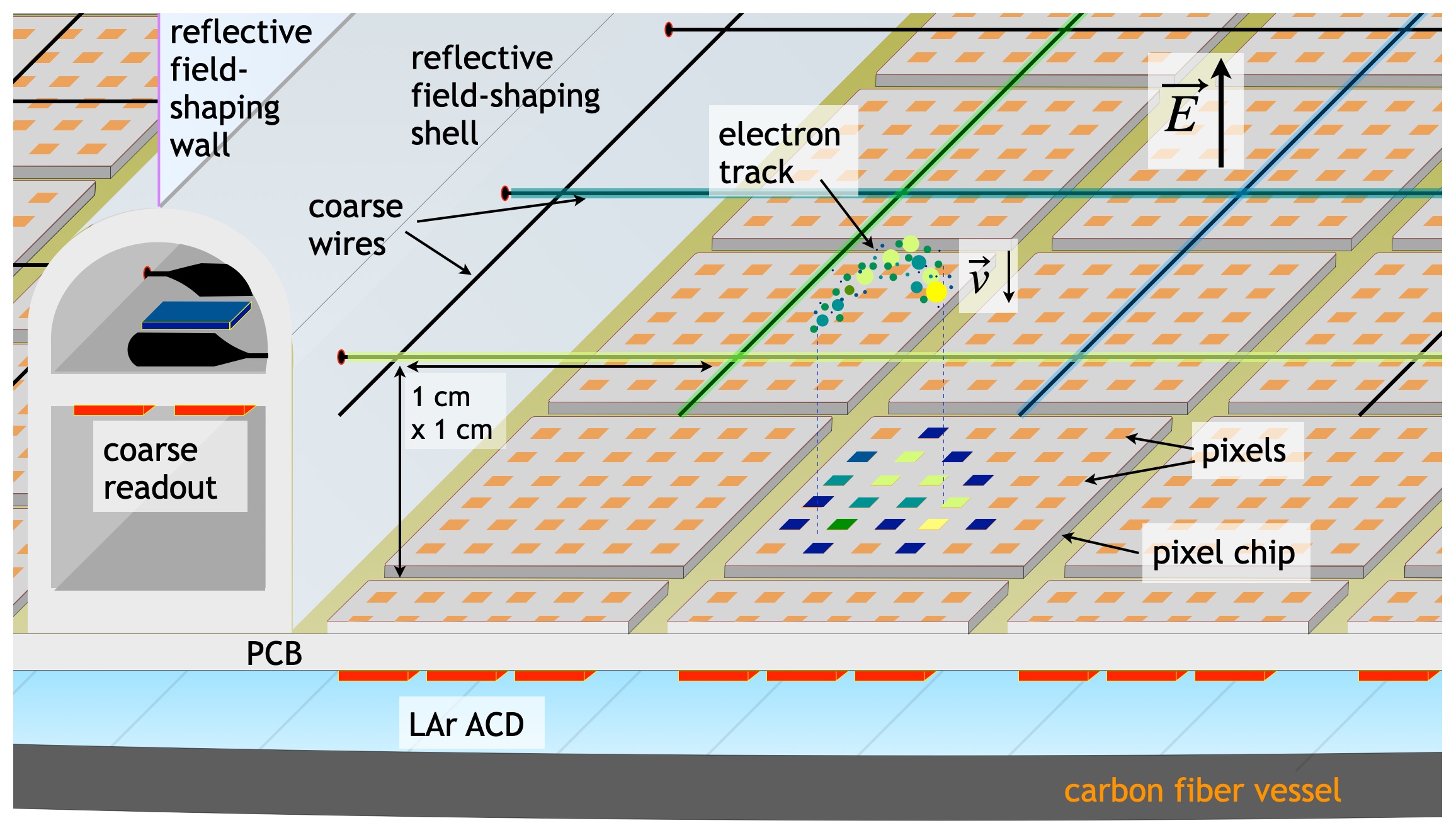}
    \caption{Schematic diagram of GAMPix implemented for GammaTPC, as described in the text.  The electron is shown just as it passes through the coarse girds, and the pixels and coarse wires with signals are highlighted with color.}
    \label{fig:GAMPix}
\end{figure}

\begin{figure}[htbp]
  \centering
  \includegraphics[width=0.48\textwidth]{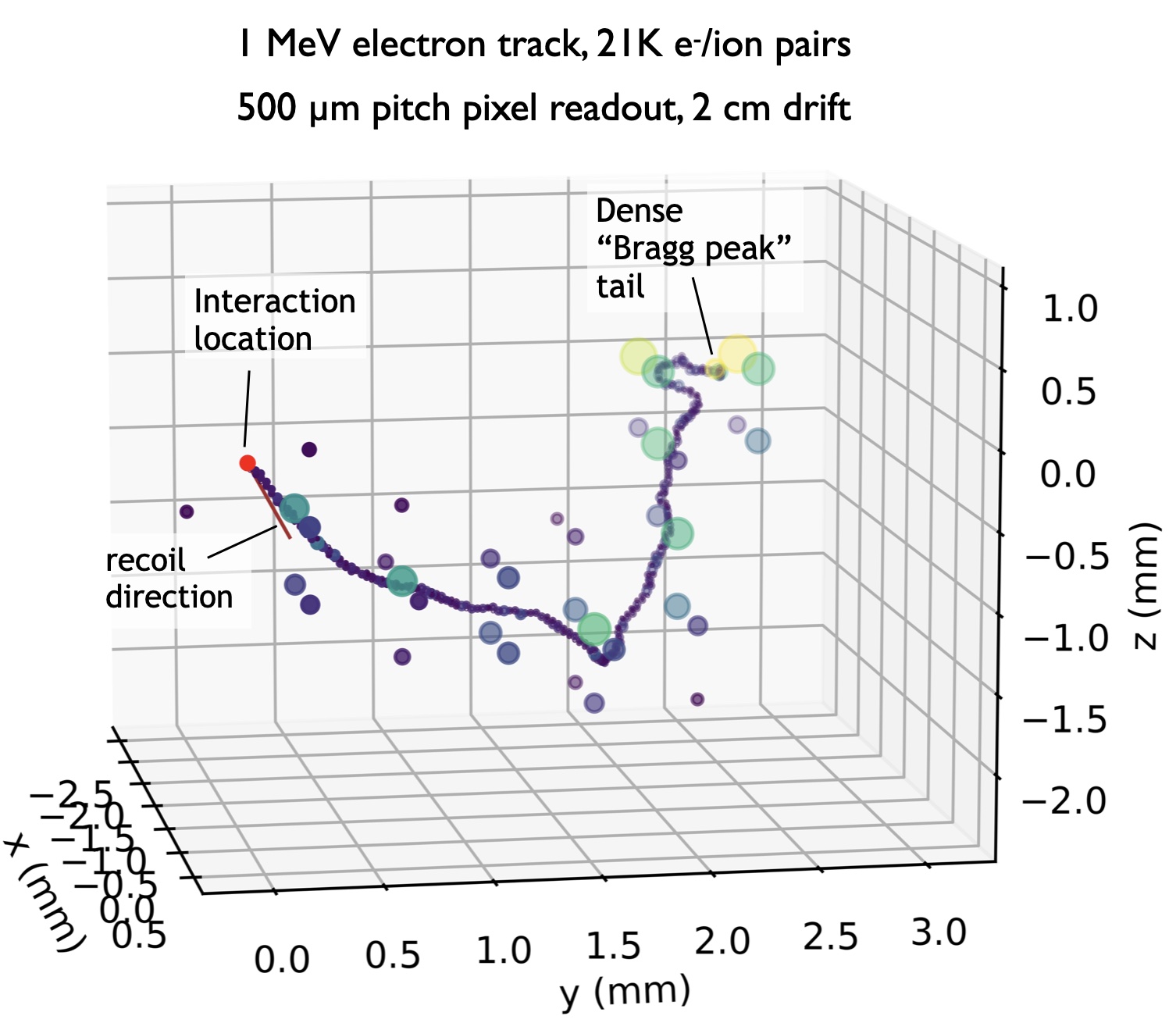}
  \includegraphics[width=0.48\textwidth]{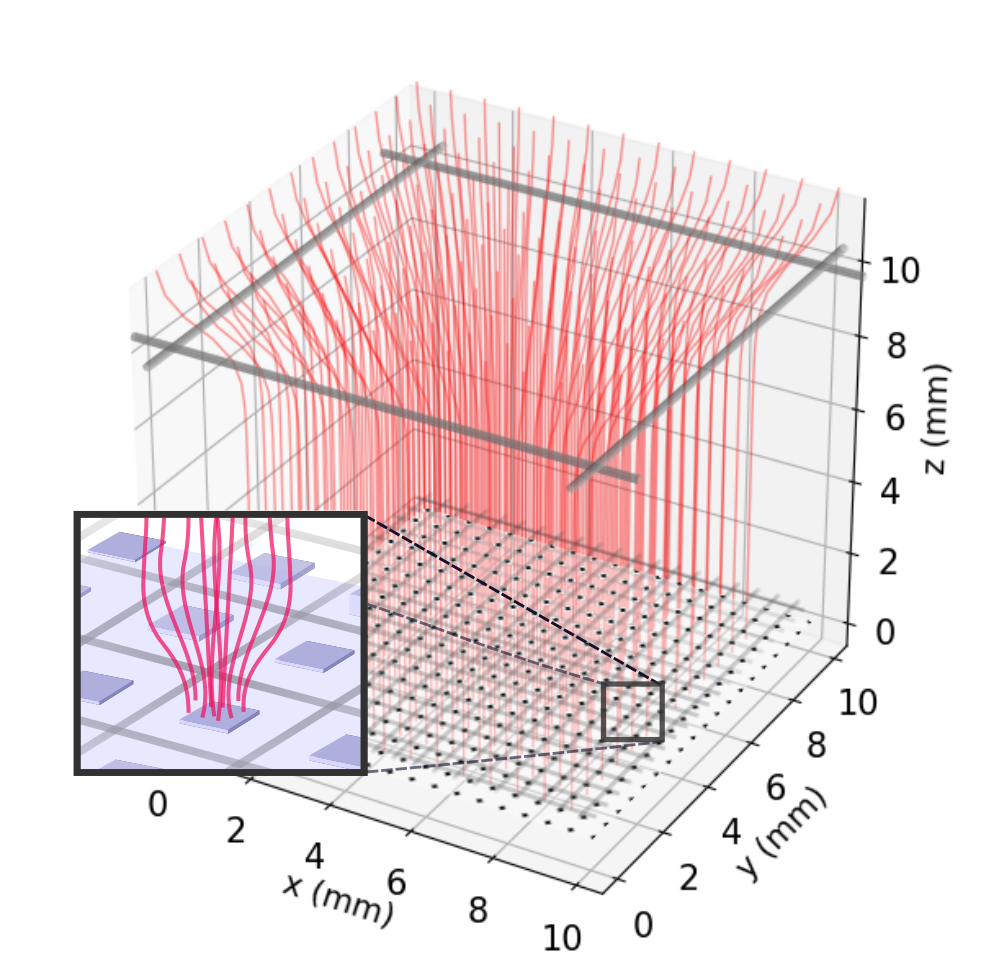}
  \caption{(left) The track from Fig.~\ref{fig:ElectronTrack}, now superimposed with pixel samples obtained with $25\,e^-$ ENC and at $2\,cm$ drift. (right) The geometry and electric fields (in red) within a single set of GAMPix wires, with a schematic inset showing of $500\,\mu m$ pitch pixels the in-plane electrode used to for the final focussing of the field onto the pixel pads.}
  \label{fig:pixels_unit_cell}
\end{figure}

The scheme we propose to solve these challenges is shown in Fig. \ref{fig:GAMPix}.  The drifting charge is measured twice, the first time with a set of coarse induction electrodes which the charges pass through. The charges are then collected on a 500\,$\mu$ m pitch array of pixel electrodes implemented directly on the cover layer of circuit-board-mounted ASIC chips. The coarse electrodes are individually-read-out crossed $x$--$y$ wires. The 1\,cm pitch is much larger than the maximum diffusion, so they are thus immune to the diffusive charge loss described in \ref{two_challenges}. They thus provide the basis for the measurement of the integral charge, while the pixels provide fine grained imaging of the track.  Note the induction signals on coarse grids are bi-polar and highly position dependent, but when combined with the pixel signals allow the charge integral to be recovered, as discussed in section~\ref{charge_integral}.  An example of the pixel image  of the track thus obtained is shown in Fig.~\ref{fig:pixels_unit_cell}~(\emph{left}).

Crucially, the coarse grids also provide a trigger signal that is used to power cycle only those pixel chips that will collect the relatively sparse set of tracks, solving the otherwise insurmountable pixel power problem. This meets the two challenges posed in section~\ref{two_challenges}.  As a bonus, the combination of the diffusion-independent coarse signals and diffusion-impacted pixel signals provides an independent measure of the drift distance as we describe in section~\ref{sec:track_imaging}. This helps overcome the limitations of pileup in a TPC. We have named this new architecture, which is a fundamental advance in the ability to image diffusion-limiting signals in a TPC, GAMPix, for Grid-Activated Multi-scale Pixel readout. 

The  electric fields around the GAMPix electrodes, set by the electrode bias voltages, are shown in Fig.~\ref{fig:pixels_unit_cell}~(\emph{right}).  Note that no drift field lines terminate on the coarse wires (which would cause that charge to be unmeasured by the pixels), a transparency condition \cite{Blum2008} which requires the field to be focused and enhanced, leaving a ``shadow'' under the wires.  We take advantage of this shadowing by aligning this with gaps between chips visible Fig. \ref{fig:GAMPix}, which provides space for wire bonding to the circuit board below them. As the charges approach the pixel plane, a set of focusing electrodes implemented in the same CMOS cover layer as the pixel sensors, and operated an expected voltage of 50-100 V, is used to focus the charges onto the pixels.  

\subsection{GAMPix in detail}\label{gampix_detail}

In Table \ref{tab:gampix_parameters} we list a number of parameters for the GAMPix architecture, of which several items bear further discussion. The requirement on noise per coarse is more stringent than the value stated in section~\ref{requirements}, as that value referred to the noise if measured \textit{with a single wire}, but now the signal is split over 4 wires and only $\sim$80\% of the signal is measured, so the noise per wire must be lower. This requires an exceptionally low capacitance per wire, which fortunately is met with the current design. The pixel noise requirement is loose because its impact is a gradual worsening of imaging.  The noise expectations are discussed in sections~\ref{pixel_asic} and \ref{coarse_readout}.

The power requirements, as noted above, derive from cryogenic considerations, and the values shown in Table \ref{tab:gampix_parameters} at this time only capture the power in the low-noise, high-power-per-channel front end amplifiers (see section~\ref{charge_readout}), which we expect to dominate the readout power.  The average pixel power is the product of the on power per pixel times the number of pixels per area, reduced by the power duty cycle for the pixels chips.  The duty cycle is the product of the number of $cm^3$ trigger voxels per event derived from the coarse gird signals, the flux of gamma rays, and the $10\,\mu s$ needed to read each $cm^3$ voxel.  We conservatively estimate this at $10^{-4}$, for which the average power requirement is comfortably met.

A key challenge to realizing this scheme is that the pixel chips must be powered up and brought to a stable state to perform a low noise measurement much faster than the $\sim\,10\,\mu s$ drift time between the coarse grids and pixels.  As we discuss in section~\ref{pixel_asic}, based on transistor level modelling of the front end amplifier, we believe this requirement can be met. One potential concern is boiling of the LAr when the chips are on, however given the low duty cycle and $<$\,100\,$\mu$s on times, we estimate that with the 100\,W/cm$^2$ power within a pixel chip, the chip temperature stays well below 1\,K heating, which should not be an issue given the expected sub-cooling we plan for GammaTPC~\cite{gammatpc24}.  The coarse grid channel count is lower by $2\times 10^4$ than the pixel count, so the power constraints on the this readout are less severe.

\begin{table}
    \centering
    \begin{tabular}{|l c | l c|}
        \hline
         \textbf{Coarse grid wires}&  &  \textbf{Pixels}& \\
         \hline\hline
         Pitch&  $1\,cm$&  Pitch& $500\,\mu m$\\
         Wire length&   $\leq\,20\,cm$&  Pad size& $200,\mu m$\\
         Wire diameter&  $100-200\,\mu m$&  Pad capacitance& $500\,fF$\\
         Capacitance&  $\leq\,2-3\,pF$&  Noise - requirement& $<\,\sim 50\,e^-$\\
         Noise - requirement&  $<\,30\,e^-$&  Noise - expected& $<\,25\,e^-$\\
         Noise - goal&  10 $e^-$&On power - expected & 0.2\,mW/ch\\
         Noise - expected&  $<\,20\,e^-$ &Average power - requirement& $\leq\,3\,W/m^2$\\
         Power - requirement&  $\leq\,1\,mW/ch$ & Average power - expected& $\leq 0.8\,W/m^2$\\
        \hline
    \end{tabular}
    \caption{Preliminary parameters of GAMPix for GammaTPC.}
    \label{tab:gampix_parameters}
\end{table}

The rib structures shown in Fig.~\ref{fig:GAMPix} that anchor the wires are fabricated of fiberglass and line the perimeter of the TPC cells. They house the coarse grid readout electronics, also anchor the reflective field-shaping walls that separate TPC cells, and must satisfy several design constraints.  First they must shape the electric fields.  An additional level of focussing must be added to the drift field, similar to that in Fig. \ref{fig:pixels_unit_cell}, but now at the full TPC cell level so that all the field lines land on pixel chips and none on the ribs. The drift field must also be fully shielded from whatever highly non-uniform fields result from the coarse readout electronics within the structure, some of which may be at the anode ground voltage.  Both of these goal are accomplished by deploying field shaping electrodes on the ribs, and tuning those and the electrodes on the walls near the ribs to properly shape the field.  The ribs must also have the same reflector and wave-shifter coating as the walls to maximize light collection.  Finally, the structure should have as small a footprint as possible in order to minimize the dead volume in the detector.  The ribs shown in Fig. \ref{fig:GAMPix}, for a baseline TPC cell dimension of 17.5 cm, occupy 1.3\% of the baseline TPC cell volume in GammaTPC. 

We close this section by considering energy deposited between the grids and pixels, which is imperfectly measured.  This is important because the anode plane is located on the outer surfaces of the detector, near where scatters are most likely to happen (note that Fig.~\ref{fig:GAMPix} shows the anode on the lower surface of the detector).  A pixel signal will be obtained for all tracks which reach the pixels after the pixel chip powers up\footnote{More precisely, a good pixel measurement is obtained if the chip powers up before the track is within a just over a sample $z$ distance which is equal to the pixel pitch.}.  With our current expected $<\,1\,\mu s$ power-cylce speed (see section~\ref{pixel_asic}), this means all tracks further $\sim\,1\,mm$ from the the pixels, which is the vast majority.  The coarse signal will be reduced for all these tracks between grids and pixels.  At a distance $z$ from the pixels and with $s$ being the distance of the coarse grids to the pixels, the coarse grid signal is reduced by the ratio $z/s$. Since the timing tells us $z$, this can be corrected, albeit with an decrease in signal to noise. This last point illustrates a key benefit of a TPC which is that in general regions with imperfect response tend to nonetheless have active material, so that there are handles in the data to at least recognize and often correct whatever error has occurred. 

\subsection{Pixel ASIC}\label{pixel_asic}

\begin{figure}
    \centering
    \includegraphics[width=0.95\textwidth]{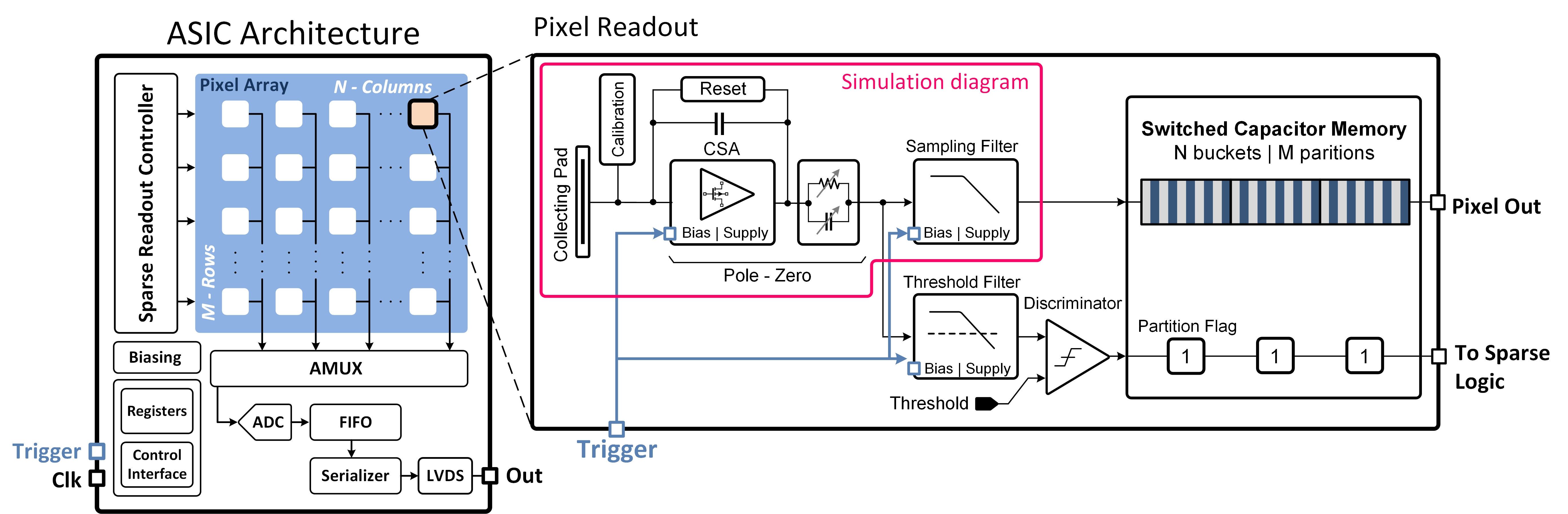}
    \caption{(left) Block diagram of the full ASIC architecture; (right) Pixel readout concept. Red area highlights the schematic diagram used for simulations.}
    \label{fig:asic_architecture}
\end{figure}

Perhaps the most demanding challenge in the development of the GAMPix readout system is the pixel ASIC, which must have $\mu$s power switching, extremely low noise in a large array of pixels, and a back end that includes digitization. To meet this challenge we have developed a preliminary architecture for the complete system-on-chip (SoC) pixel ASIC, shown in Fig.~\ref{fig:asic_architecture}. This configuration encompasses a pixel readout system incorporating a charge-sensitive amplifier (CSA), filter stages, buffered switched capacitor analog memory, and per-pixel triggering. Following this, the front-end analog signals are multiplexed and digitized to facilitate digital data transmission out of the chip for further analysis. Notably, integrating a power pulsing mechanism, driven by the trigger signal, will significantly reduce the overall power consumption of the ASIC. Leveraging the low duty cycle of each chip, we expect to meet the demanding W/m$^2$ power requirement, while efficiently generating highly sparsified data. This work directly builds upon the design of low-noise pixelated ASICs for x-ray detectors \cite{epix,epix10k,sparkpix} and the CRYO ASIC design at SLAC~\cite{CryoAsic1, CryoAsic2, CryoAsic3}, a SoC multi-channel (non-power cycling) charge readout developed for the nEXO experiment~\cite{nexo_2018}, and fits into the broader context of the development of several cryogenic ASIC readout systems for DUNE ~\cite{CryoAsic1, Dwyer_2018, Adams_2020}.

The pixel readout of Fig.~\ref{fig:asic_architecture}~(\emph{right}) highlights in red the functional block diagram used for initial simulations to assess the noise and power cycling characteristics of the CSA. We have explored two two potential low-power CSA circuit architectures, denoted here as CSA-1 and CSA-2. CSA-1 follows a conventional approach employing a single-stage amplifier with a PMOS input device and a cascode load~\cite{Rivetti} to achieve high DC gains in open loop. In contrast, CSA-2 adopts an inverter-based configuration commonly found in A-to-D converters such as sigma-delta modulators for audio applications~\cite{Luo2013}. For this design, the amplifier leverages self-cascode structures~\cite{DXu} to further enhance the DC gain, surpassing 80\,dB with the given technology. Achieving a high DC gain is essential for improving the virtual ground of the circuit, particularly when configured in a closed loop with a feedback capacitor. The pole-zero cancellation currently uses passive components with the amplification factor chosen to ensure an output swing within the supply limits. An active CMOS realization of this stage is currently under development and will follow previous solutions~\cite{Geronimo_2000, Rivetti}. The sampling filter is modeled at the behavioral level based on the CRYO ASIC design and features a fifth-order Bessel topology with a shaping time of 0.6\,$\mu$s.

\begin{figure}[t]
    \centering
    \includegraphics[width=1.0\textwidth]{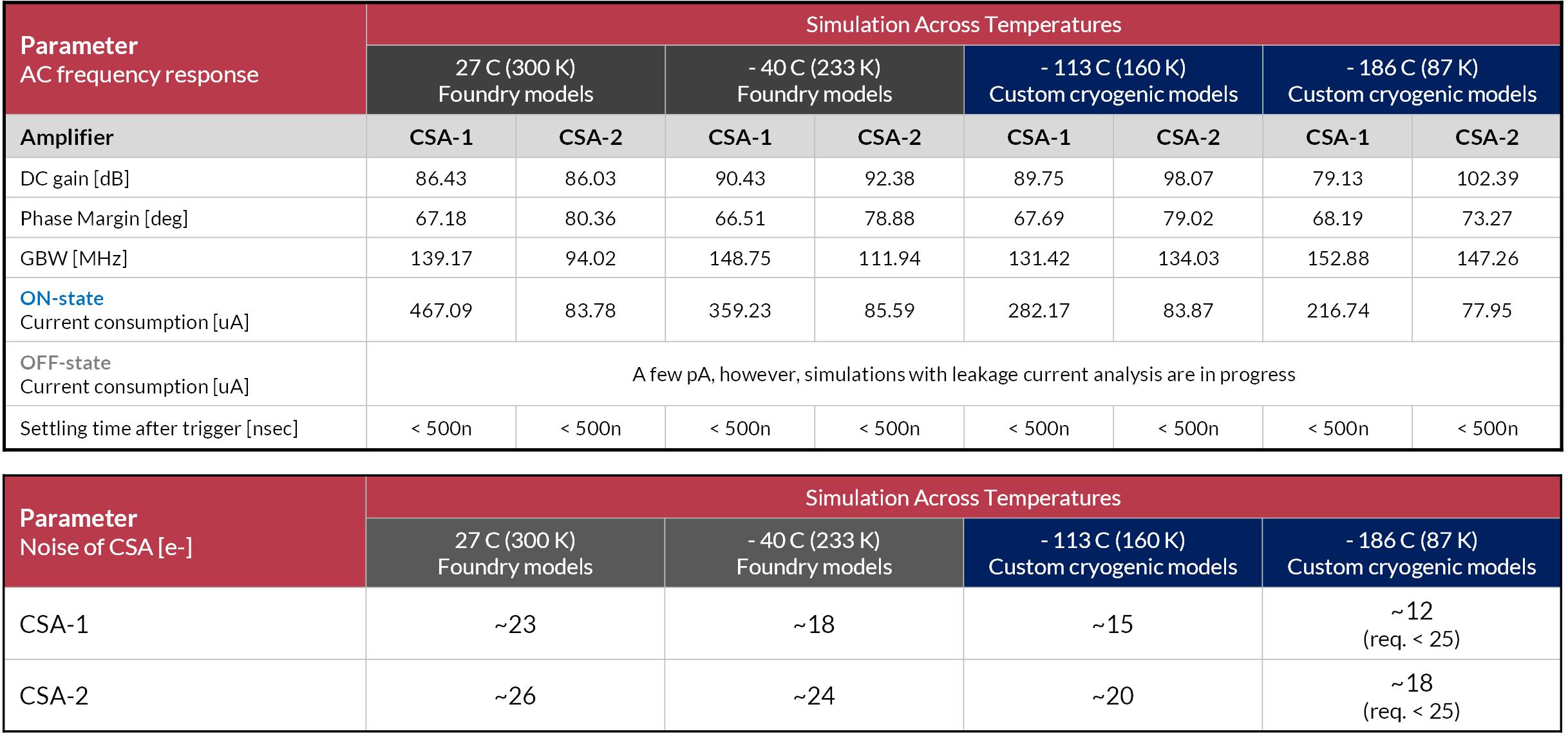}
    \caption{Preliminary simulations across temperatures using the functional block diagram of Fig.~\ref{fig:asic_architecture}\,(right) with the two CSA circuit solutions. (Top table) Performance summary from the AC loop stability and DC analysis. (Bottom table) Estimated noise.}
    \label{fig:pre_performance}
\end{figure}

Preliminary simulations were initially conducted using models provided by the foundry, which are suitable for the industry-standard temperature range (i.e., 27$^{\circ}$C to $-$40$^{\circ}$C). Subsequently, custom cryogenic models were utilized to extend the simulations to LAr conditions. These cryogenic models were developed at SLAC based on measurements of 130\,nm CMOS process test structures at 160\,K and 87\,K using a dedicated liquid nitrogen-based test bench system. The models incorporate the same device portfolio as the foundry models and have been validated with the CRYO ASIC. For the charge readout of this work, we have selected nominal and low-vt 2.5\,V devices for the front-end pixel design which will be incorporated in the fast power switching scheme to minimize the leakage current in the off state and fit within the overall $\sim$\,1\,W/m$^2$ power budget. Fig.~\ref{fig:pre_performance}~(\emph{top}) presents a summary of the CSA performance across temperatures for the two previously described CSA architectures. The results include the loop stability analysis and the settling time of the circuits when the trigger signal is issued. In all cases, the resulting DC gain exceeds $70$\,dB and the phase margin is over $65$ degrees. A power pulsing mechanism is employed in the amplifier through the trigger signal, directly impacting the main bias circuitry to halt the currents flowing through the circuit branches. This approach minimizes overall current consumption. When triggered, the circuit is activated, achieving a settling time of less than $500$\,ns across temperature conditions. This rapid response is specific to the CSA, and additional timing analysis will be conducted with the pixel readout signal chain once the extra circuits are fully implemented at the transistor level.

The preliminary readout noise of the CSA for the anticipated 500\,f\/F capacitance pixel pads, maximum charge per pixel of 5.5\,k$e^-$ and the given shaping time, is summarized in Fig.~\ref{fig:pre_performance}, \emph{(bottom)}. Note that so far the noise only includes the contribution from the CSA and not yet the surrounding components and Bessel filter. The initial on-state current consumption for CSA-1 is less than $300\,\mu A$, while for CSA-2, it is below $90\,\mu$A in cold environments. This translates to approximately $600\,\mu W$ and $180\,\mu W$, respectively with a supply voltage of $2.0$\,V. In the off-state, the current consumption is reduced to a few pA. However, it is crucial to acknowledge that this result may be overly optimistic as it does not fully account for leakage current analysis, which is currently underway. Additionally, the consumed current of both CSA circuits is subject to further reduction through proper design modifications.

The results reported above are promising, as the CSA represents the core of the pixel and is the primary contributor to noise and power consumption in the signal chain. Extensive simulations and noise analysis will be conducted while we optimize the pixel design and transition to an implementation fully in CMOS.

\begin{figure}[t]
    \centering
    \includegraphics[width=1.0\textwidth]{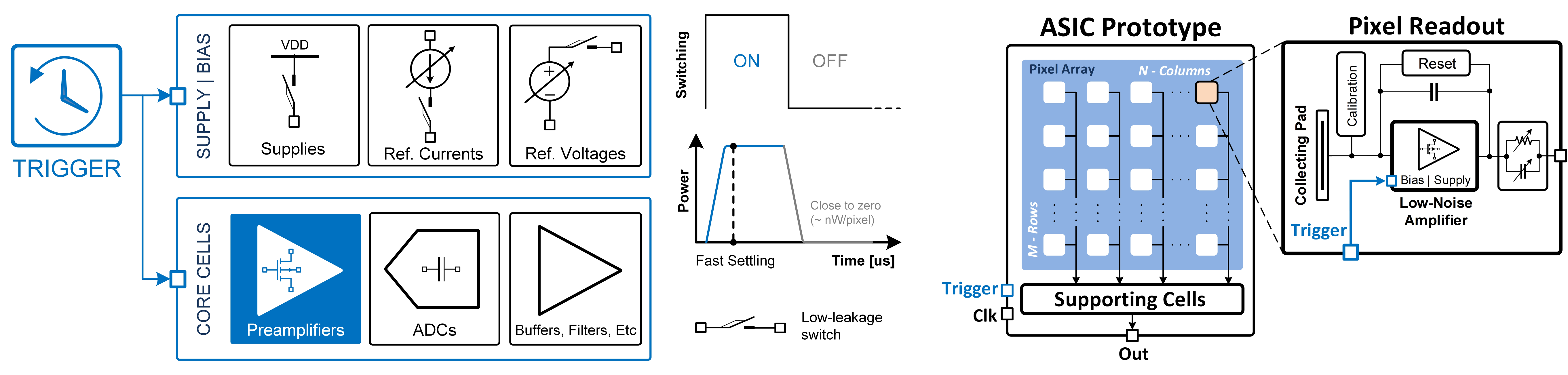}
    \caption{(left) Concept of power pulsing. (right) reduced functionality R\&D prototype ASIC.}
    \label{fig:switching_small_asic}
\end{figure}

One of the key challenges of the GAMPix approach is power cycling. To minimize average power consumption (our initial goal is $\lesssim5$\,W/m$^2$) the ASICs operate in a power-down mode until activated by the coarse grid signal. When triggered, an ASIC needs to reach a stable operational state in the electron drift time between the coarse grids and the pixels (5--10\,$\mu$s for this proposal). This led us to three major design considerations. First, the power supply needs to deliver the peak currents required for the fast turn-on. Second, once ramped up, supply voltages need to be stable at their operational values to avoid introducing extra noise into critical analog sections. Finally, the off-power dissipation must be essentially zero ($\lesssim\,200$\,nW/pixel as initial goal), requiring careful design of the pixel power system. 

Our approach to addressing these requirements is shown in Fig.~\ref{fig:switching_small_asic}\,\emph{(left)}, where a switching mechanism directly drives the supplies and the bias of core active cells. The concept is mainly applied to the pixel preamplifier, but it can be extended to other power-hungry cells of the ASIC (i.e., ADCs, analog buffers, filters, etc.). The pixel readout, analog, and digital sections of the chip will be powered independently to minimize potential noise coupling between them. Nominal and low-vt 2.5V devices will be used for the switching implementation to keep residual currents during the off state at minimum. The design will leverage SLAC expertise implementing power pulsing techniques on previous projects such as kPix\,\cite{Kpix}. In terms of the ASIC design, we aim to produce a chip prototype with several pixels, sufficient for a proof-of-principle demonstration (Fig.~\ref{fig:switching_small_asic}\,\emph{(right))}. This is indeed crucial to retire the two most challenging and highest risk parts of the design as soon as possible: \emph{power switching and noise}. The targeted size of the prototype is $\sim$5$\times$5\,mm$^2$ and it will be integrated with a common region of supporting electronics with reduced functions (i.e., programming the device, only preamplifier readout, etc.). The front-end power-switched pixels will have the full functionality necessary to demonstrate the principle. 

\subsection{Coarse readout and trigger}\label{coarse_readout}

The primary requirement for the coarse wire readout is the noise level shown in Table \ref{tab:gampix_parameters}. This has been demonstrated with a CMOS ASIC operating at 77\,K, resulting in a $20\,e^-$ ENC with a 3\,pF input capacitance, and 12\,$\mu$s shaping time \cite{Deng_2018}, serving as our benchmark. Initial simulations have been conducted utilizing the CSA architectures outlined in Section \ref{pixel_asic}, with the input device tailored to the specified load capacitance. Preliminary results indicate an ENC of approximately $30\,e^-$ with a 3.6\,$\mu$s peaking time on the Shaper filter. However, this outcome is not conclusive as there is a margin for improvement in the preamplifier stage, and trade-offs involving current consumption and longer peaking times are being explored to achieve the desirable noise levels.

To minimize stray capacitance, each coarse wire or possibly pair of coarse wires will have separate readout chip connected directly to the wire ends. These will likely be operated at the grid voltage of roughly $-500\,V$ relative to the pixels, though AC coupling to grounded readout could be considered.  The analog signals from all the wires will be routed to a single DAQ and trigger chip per TPC cell. The trigger derived from these signals is discussed in section~\ref{triggering} below.

Due to the stringent noise specifications, we anticipate encountering a dynamic range challenge given the wide range of energies that GammaTPC will measure.  This could be addressed with a dual gain CSA design, a strategy commonly employed in x-ray ASIC detectors to prevent signal saturation \cite{epix,epix10k,sparkpix}, or conceivably by an additional set of $x$--$y$ wires with a low gain readout. The requirements for this have not yet been fully developed, and will likely be driven by the acceptance we will require for those high energies with long straight tracks that are close to vertical.

\section{Integral charge measurement}\label{charge_integral}

Here we discuss how the integral charge is measured from the combination of coarse grid and pixel signals.  As we noted in section~\ref{gampix_overview}, the coarse grid signals are bipolar and highly position dependent. Nevertheless, it's crucial to obtain an accurate measure of the charge integral by using the wire grid. While a combination of wire signal and pixel imaging data should yield a good result, initially it wasn't obvious whether this could be achieved with the necessary accuracy.  Achieving this is the second significant challenge of the GAMPix development project.  

To complicate matters, the situation becomes even more complex when considering a set of tracks, though distinct, being close enough in space to induce overlapping signals on the same set of coarse wires. Simulations showed that this occurs for a non-negligble fraction of events, and so must be handled. We approached this problem by first building a high fidelity simulation of the complete chain of signal generation on both the wires and pixels. We then developed a robust recursive procedure using this signal generator that recovers the charge integral to high precision even in a crowded field of tracks.

\subsection{Simulation method and software tools}\label{simulation_method}


A precise model of the GAMPix readout geometry in a TPC cell has been developed using COMSOL - a tool for calculating electric fields in 3D geometries. The model includes the biased wire grid, pixel plane and appropriate boundary conditions. By using an electric field of $0.5$ kV/cm, we drift electron tracks in liquid Argon between the anode and the cathode. 

Based on Ramo's theorem~\cite{Ramo_39}, the signal \( s(t) \) induced on an element within the TPC by a charge moving with velocity \( \vec{v} \) is directly proportional to the velocity of the charge and the gradient of the weighting potential \( \phi \). The weighting potential \( \phi \) is defined as the potential that would be produced at a given point in space if the electrode under consideration were held at unit potential and all other conductors were grounded. 

$$
s(t) = q\vec{v} \cdot \,\vec{\nabla}\phi 
$$

\begin{figure}[htbp]
  \centering
  \includegraphics[width=0.95\textwidth]{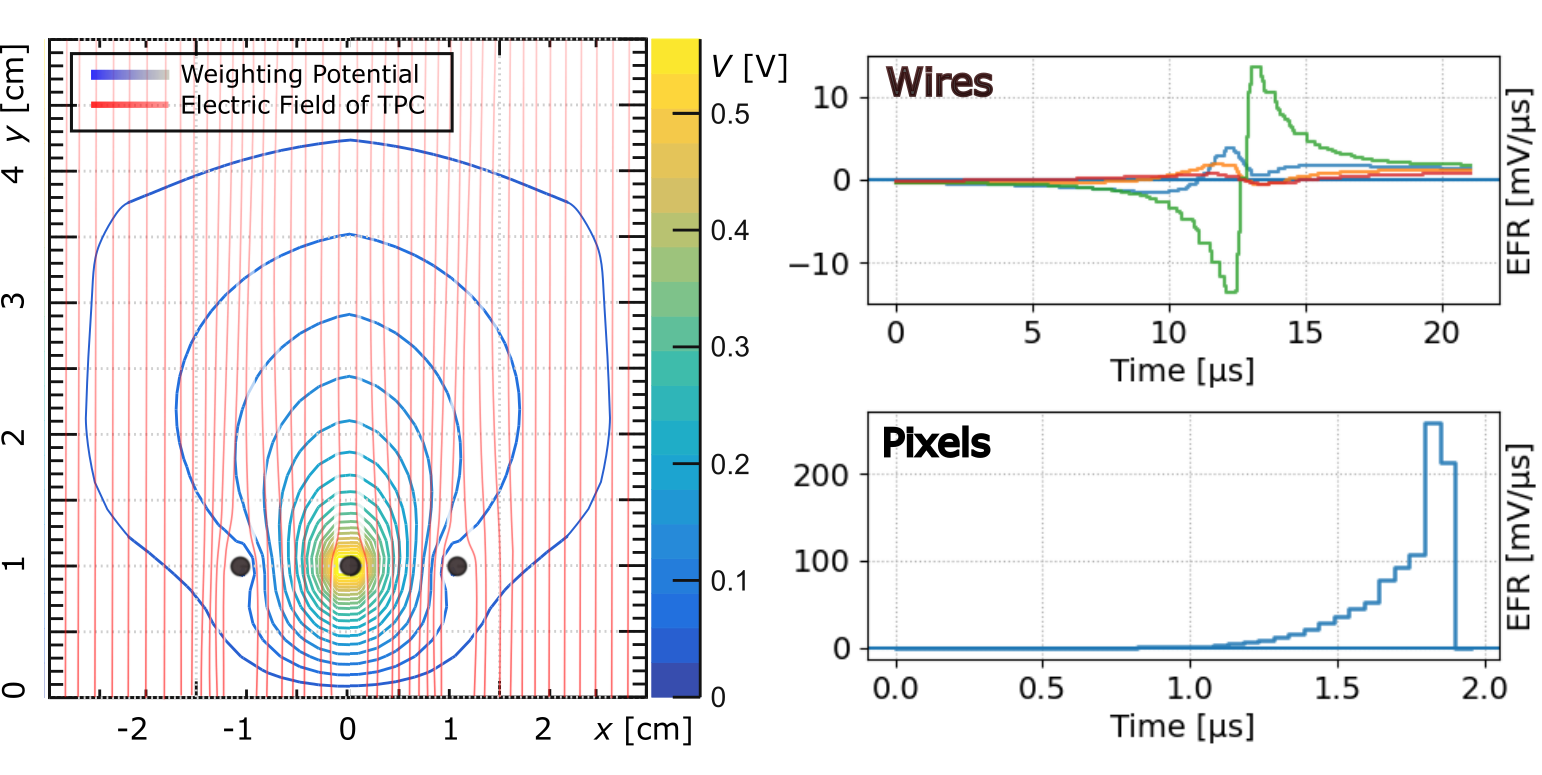}
  \caption{\textit{Left:} Illustration of the overall electric field in the cross-section of the GAMPix (in red) and the weighting potential of the central wire (blue-yellow). The weighting potential is used to calculate the Electric Field Response (EFR) function of sensors. 
  \textit{Right:} EFR of Wires and Pixels. The EFR of Wires depends on the location of the charge relative to the wire, whereas the EFR of the pixels is position independent. }
  \label{fig:garfield}
\end{figure}

The weighting potential of a wire, as illustrated in Fig.\ref{fig:garfield}, is utilized by Garfield++ to compute the signal on that wire induced by moving charges. For each sensor in  TPC designed to detect signals, we simulated an Electric Field Response (EFR)  function. This function represents the signal generated by a unit charge moving at unit velocity along a specific electric field line, for various drift paths of these unit charges: 
$$EFR_{x,y} = q_e\vec{v}_{||\vec{E},xy} \cdot \,\vec{\nabla}\phi $$
where $q_e$ is a unity charge and $\vec{v}_{||\vec{E},xy}$ is a unity velocity along the $\vec{E}$ electric field at position (x,y). A track drifting at position (x,y) will produce a signal equal to $$ s(t) = qv_{(t)}EFR_{x,y} $$

 As shown in Fig.\ref{fig:garfield} the shapes of the response functions (EFRs) of wires and pixels are quite different. When charges are approaching the wire grid, they induce a faint negative signal on the wires. At a distance of around one wire pitch, the signal reaches its minimum, with the amplitude proportional to the total charge of the track. As charges pass the wire grid, the induced signal becomes positive and decays as the charges approach the pixel plane. By contrast, the response function of the pixels is similar to a rising exponential up to the point where the charges reach the pixel. The signal is unipolar and happens on a much shorter timescale than the wire response function. 

The linearity property of adding induced signals facilitates the simulation of complex track signals. Conceptualizing a drifting track as an electrical current, we spatially bin the charge distribution and adjust the z-dimension according to the drift velocity, resulting in a current $I(x,y,t)$ that varies with both position and time. The final induced signal is then derived by convolving this current with the spatially dependent EFR function across the time domain.

\begin{figure}[ht]
  \centering
  \includegraphics[width=0.92\textwidth]{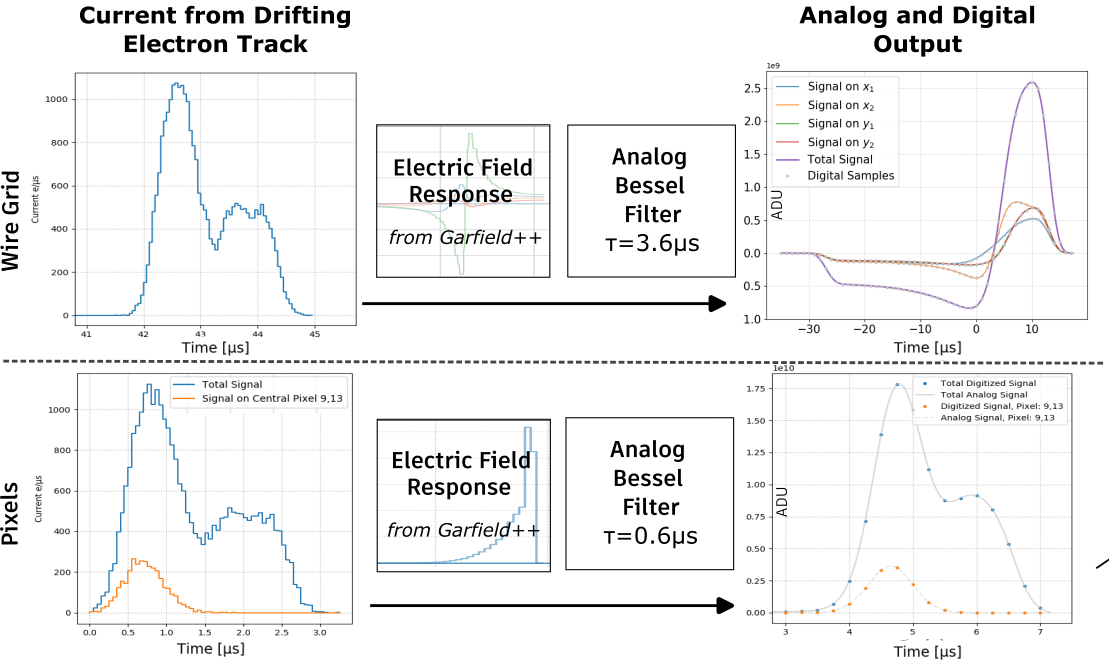}
  \caption{Signal chains in GAMPix, shown with an example track with extended structure. Each electron track is converted into a spatial current and convolved with the appropriate EFR function. The output of the wire grid is four signals (and their sum). The output of the pixel plane are signals from activated pixels (total signal in blue and signal on one pixel in orange)}
  \label{fig:signal_chain}
\end{figure}

In our setup, the front-end Charge Sensitive Amplifier integrates signals on a capacitor, followed by pole-zero cancellation that differentiates this signal, yielding an amplified current.\ref{pixel_asic} In our simulation, the current passes through a 5th order Bessel Filter with a 3.6 $\mu$s peaking time. For wire signals, this peaking time aligns with the bipolar signal's timescale, optimizing the signal-to-noise ratio without averaging out the bipolar signal. Similarly, pixel signals undergo the same filtering with a 0.6 $\mu$s peaking time, matching the pixel response function's timescale. The hardware implementation may utilize a similar filter to more efficiently mitigate noise. 

Fig.~\ref{fig:signal_chain} displays an example of the resulting signals produced by an electron track with extended structure. The wire's analog output shows bipolar signals, not resembling the electron track's current profile, yet their sum of amplitudes directly correlate with the track's total charge. Conversely, the pixel's analog output closely resembles the electron track's current profile, allowing for the 3D reconstruction of the electron track with a resolution equal to the pixel pitch (500 $\mu$m) using the digitized pixel signals.


\subsection{Triggering Mechanisms in GAMPix}\label{triggering}

The complicated nature of the coarse grid signals (Fig.~\ref{fig:signal_chain}~(\emph{upper right})) presents a challenge for using them to trigger the pixel chips.  In our simulations, we have currently implemented two schemes that will presumably form the basis for the eventual hardware implementation.  The first works for most signals, while the second achieves the lowest possible threshold, but is unreliable to power up the pixels in time with our current grid to pixel distance. 

The basis of these schemes is the bipolar shape of the wire signals generated by a track. As the track approaches the wire grid, it induces a negative signal. When the track passes through the wire grid, the signal turns positive, with a larger amplitude compared to the negative part. The two triggering schemes are:

\begin{enumerate}
    \item \textbf{Two-Stage Triggering Scheme:}
    \begin{enumerate}
        \item \textit{Priming Stage:} In this initial phase, the triggering process is primed when the aggregate signal from two neighboring wires falls beneath a predetermined threshold. This stage is crucial for preparing the system for potential track detection.
        \item \textit{Activation Stage:} Trigger activation occurs when the signal transitions to a positive value. Subsequently, a calculated delay of approximately $5\,\mu$s is introduced, after which the trigger activates the pixels located directly beneath the active wire pairs. This method is effective at appropriately timing the powering of the pixels, and it's primarily effective for detecting tracks that fall within the pixel's sensitivity range.
    \end{enumerate}

    \item \textbf{Positive Threshold Triggering Scheme:}

    In the case when the negative wire signal fails to prime the Two-Stage trigger, this scheme is used. When the cumulative signal from wire pairs crosses this positive threshold, the system records the event. This scheme is particularly crucial for identifying electron tracks comprising 200 to 400 electrons, which are not effectively captured by the first scheme. Due to the relatively small size of these tracks and their low probability of pixel detection, this scheme focuses solely on event recording and subsequent extraction of the total charge from the wire signal, bypassing pixel activation.
\end{enumerate}

When it comes to complex event topologies within GAMPix, there is a potential for 'ghost' triggers, analogous to the ghost images depicted in Fig. \ref{fig:Ghosts}. This phenomenon is estimated to amplify the count of triggered pixels by a factor of $5-10$. This increase is accounted for in the projections presented in Table \ref{tab:gampix_parameters}, and currently does not seem to be a significant problem. If necessary, however, a third wire plane, angled at $45^{\circ}$ to the $x$--$y$ grid could be used and would substantially mitigate this effect. However, this would entail additional complexity and result in the distortion of fields in a pattern misaligned with the $x$--$y$ pixel grid.

\subsection{Electron Track Reconstruction}\label{track_reconstruction}

\begin{figure}[ht]
  \centering
  \includegraphics[width=0.96\textwidth]{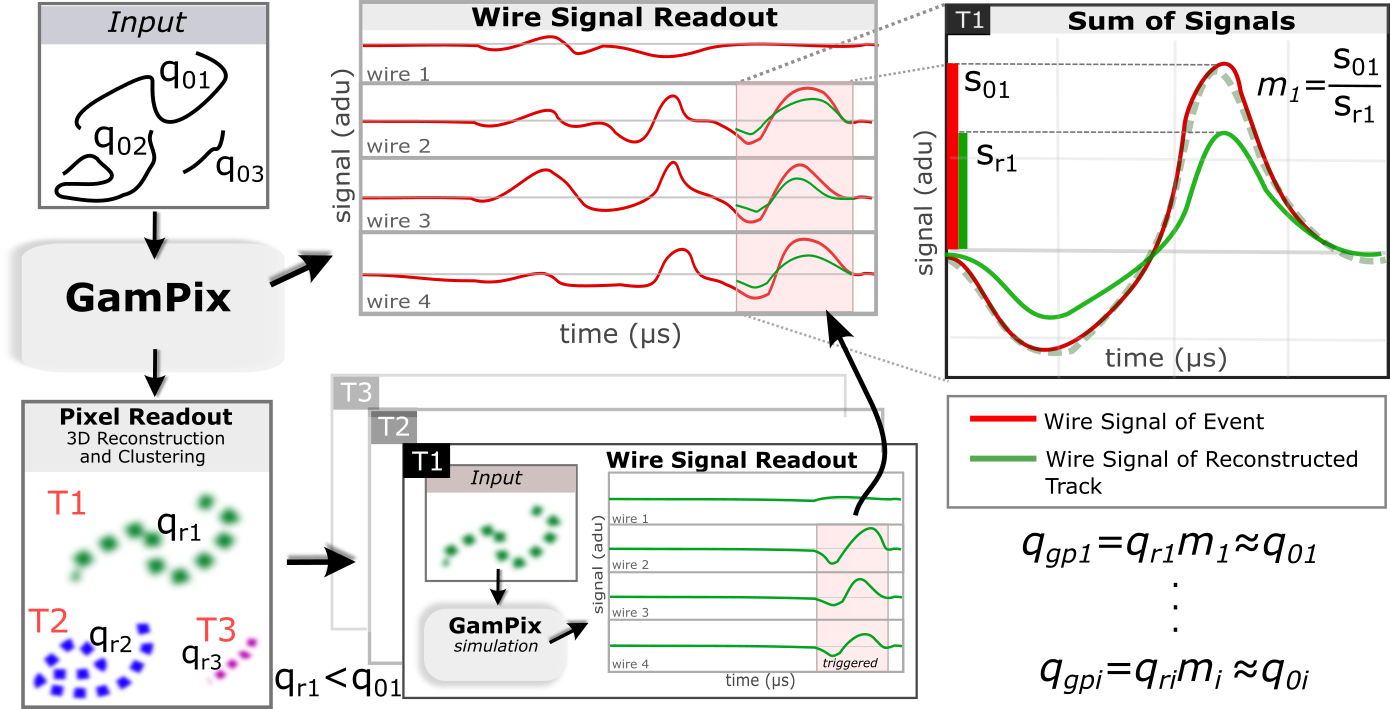}
  \caption{Example of the charge reconstruction process: A simulated event with three tracks ($q_01$, $q_02$ and $q_03$) as \textit{Input}. The \textit{GamPix} detector (currently in form of a simulation, can be hardware too) outputs wire and pixel signals. Pixels are used to reconstruct the shape of the tracks. Then, a clustering algorithm is used to distinguish the tracks. To recover the charge of the tracks, each detected track is individually rerun through the \textit{GamPix Simulation} (simulated response of GAMPix), where it is used as input. The output of that simulation enables us to a) identify the activated wires and relevant time frame and b) produce simulated wire signals. The next step is to sum up the signals from activated wires from the original event (red curves) in the relevant timeframe and compare it to the sum of the simulated track's wire signals (green curves). The simulated wire signal sum is a fraction of the original sum, but the signal's shape is almost identical. The ratio of the two signals yields the fraction of the lost charge on the pixels. Therefore, the total charge for each track is the product of the total charge acquired by the pixels and the ratio of the sums of wire signals of the original event and the simulated track.}
  \label{fig:reconstruction_algo}
\end{figure}

We now are turn to the method we developed to accurately recover the total charge and shape of simulated electron tracks. Due to the complex nature of the wire signals, there isn't a straightforward analysis that could extract the total charge of the tracks. Rather, based on our capability to accurately simulate the GAMPix detector described in section~\ref{simulation_method}, we proceed with a method sketched in Fig~\ref{fig:reconstruction_algo}. Following an event (a collection of tracks generated by PENELOPE), the pixels are triggered and data is saved into a buffer. The first step is to reconstruct the tracks from the available pixel data, but these tracks will be missing charge due to diffusive loss. To recover the total charge of tracks, each reconstructed track is individually rerun in the simulation. A clustering algorithm is used to distinguish tracks if the event had multiple energy deposition sites. The repeated run of the tracks in the simulation enables us to 1) identify the activated wires and relevant time frame and 2) produce simulated wire signals. We sum up the signals from activated wires from the original event in the relevant timeframe and compare it to the sum of the simulated track's wire signals. Since the pixels receive less charge than the track's charge, the simulated wire signal is a fraction of the original sum. However, the signal's shape is almost identical. The ratio of the two signals yields the fraction of the lost charge on the pixels. Therefore, the total charge for each track is the product of the total charge acquired by the pixels and the ratio of the sums of wire signals of the original event and the simulated track. A detailed mathematical treatment of this process is provided in Appendix \ref{appendix:first}

\subsection{Resolution on the Measure of Integral Charge}\label{integral_resolution}

\begin{figure}[htbp]
  \centering
\includegraphics[width=0.48\textwidth]{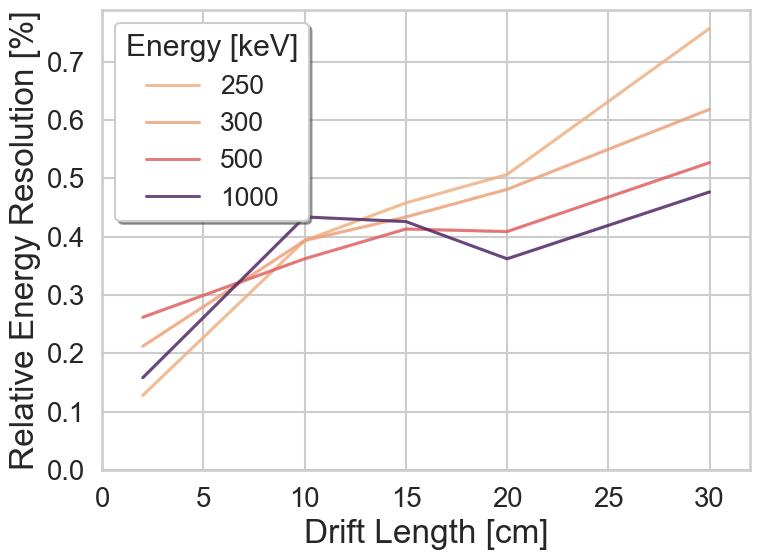}
\includegraphics[width=0.48\textwidth]{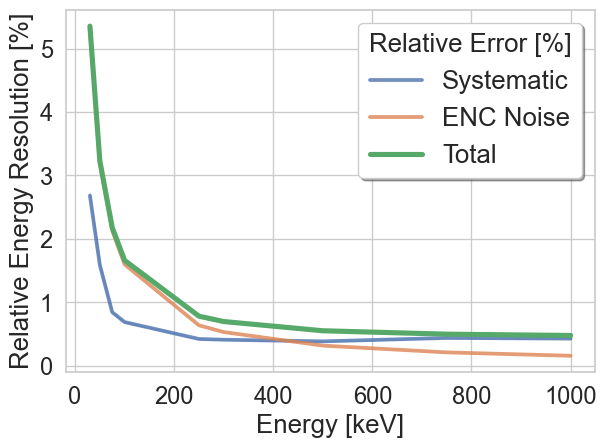}
\caption{Resolution on the integral charge of tracks measured by the GAMPix system. Mayor sources of uncertainty are the Systematic error and Front-end Noise. The constant absolute noise contribution from the front-end ENC dominates as the source of error, especially at low energies.}
\label{fig:final_resolution}
\end{figure}

The results of the technique described in section~\ref{track_reconstruction} are shown in Fig.~\ref{fig:final_resolution}.  On the (\emph{left}) we see that the systematic error is below 1$\%$ for energies greater than 250 keV and over all drift distances. Not shown are tracks of less than 50 keV and long drift lengths which have little pixel signal and have systematic errors significantly larger than 1$\%$.  Tracks of less than 25 keV have a low probability of being detected on the pixels, which prohibits us from using the aforementioned reconstruction method. In that case, we can estimate the position of the charge by looking at the difference of the signals at the wires and the shape of the signal. Due to the noise in the wires, that estimate is bound to have an error of at least 20$\%$, yielding a systematic error in energy estimation of $\sim 10 \%$. 

The final resolution as a function of deposited energy is shown on the (\emph{right}).  Despite the systematic reconstruction issues for small tracks, the electronics readout ENC is far larger for all but the highest energies, where the systematic error begins to contribute a significant fraction.  These results are very encouraging, and are at the level needed for GammaTPC.

\section{Track imaging results}
\label{sec:track_imaging}

The driving motivation behind the development of the GAMPix architecture was to obtain fine grained 3D imaging of Gamma ray interactions. In the Compton regime, this means using the pixel data to measure the interaction locations at the heads of electron recoil tracks, and also to measure those track's initial directions.  It would appear from visual inspection of the pixel samples for the track in Fig.~\ref{fig:pixels_unit_cell} that both of these should be possible. We quantified this in a machine learning (ML) study published separately~\cite{gammatpc_ml_ai, gammatpc_ml_apj}, and whose main results we present here. The approach was 3D convolutional neural networks that analyzed entire electron tracks at once, with one network producing a prediction for the track head location, and another producing a prediction for the initial scattering direction. 

\begin{figure}
    \centering
        \includegraphics[width=0.82\textwidth]{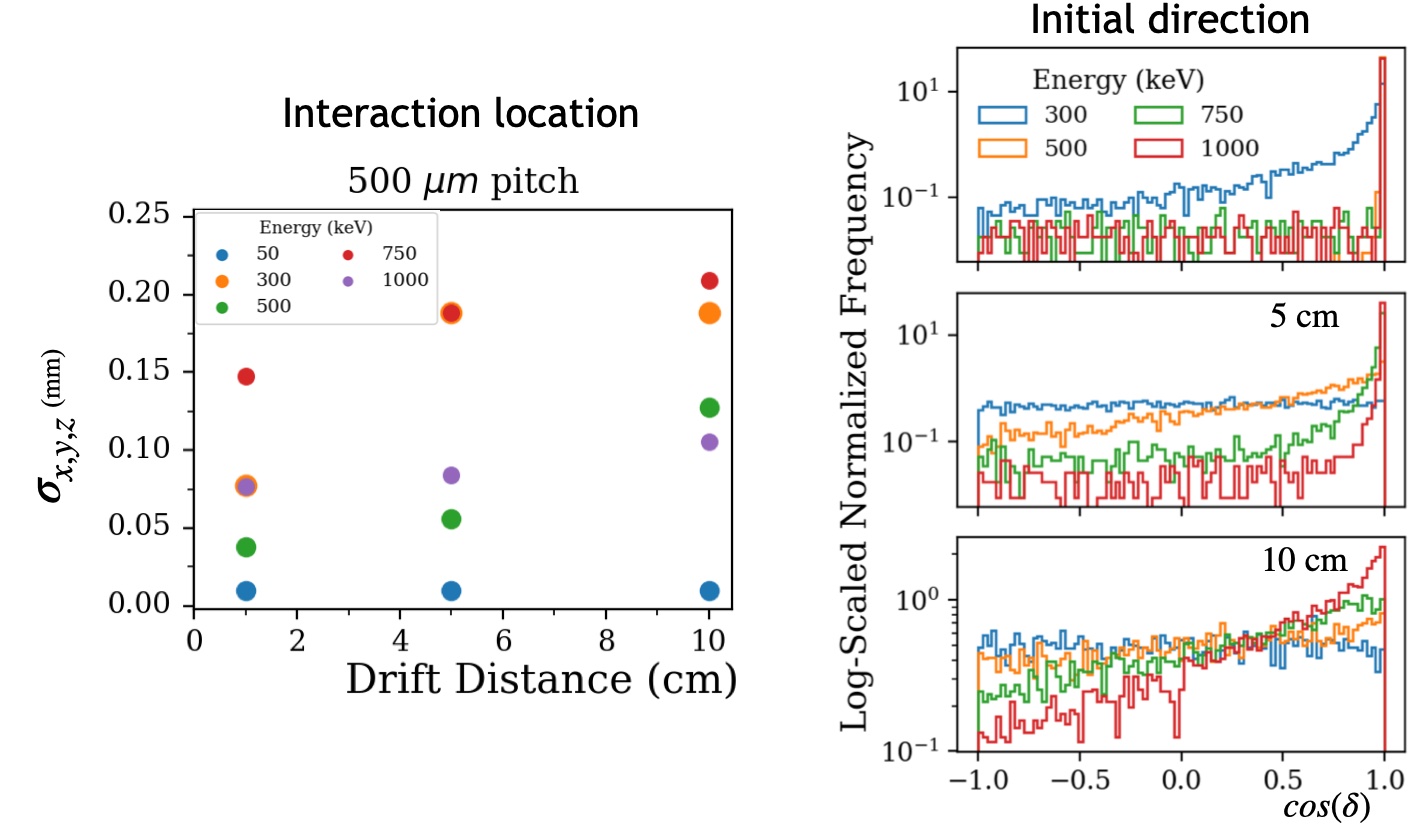}
    \caption{Track reconstruction using ML techniques described in the text as a function of track energy and drift distance, here with the baseline pixel readout noise of 25\,e-. (left) RMS resolution of the interaction location, and (right) distribution of the cosine of the angle between the true and measured initial track directions at several drift distances.}
    \label{fig:track_imaging}
\end{figure}

The results are shown in Fig.~\ref{fig:track_imaging}.  These results are not very sensitive to the pixel pitch, and that the baseline 500\,$\mu$m pixel pitch, which was motivated largely by the expected ASIC layout constraints, is a reasonable choice.  The level of position resolution is very encouraging and should be a powerful tool for the Compton telescope technique.  Even more impressive is the initial direction measurement, which is especially powerful for short drift distances.

\begin{figure}[hb]
  \centering
  \includegraphics[width=0.495\textwidth]{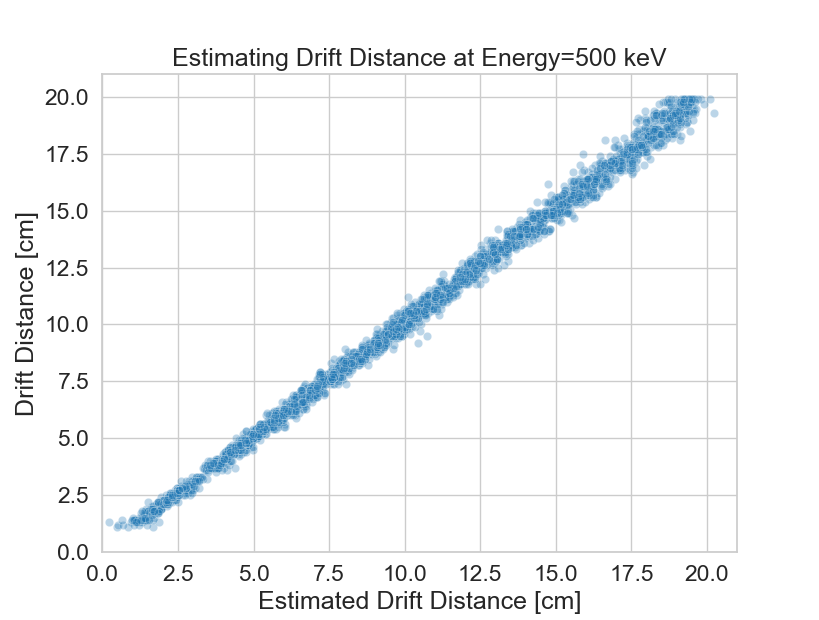}
  \includegraphics[width=0.459\textwidth]{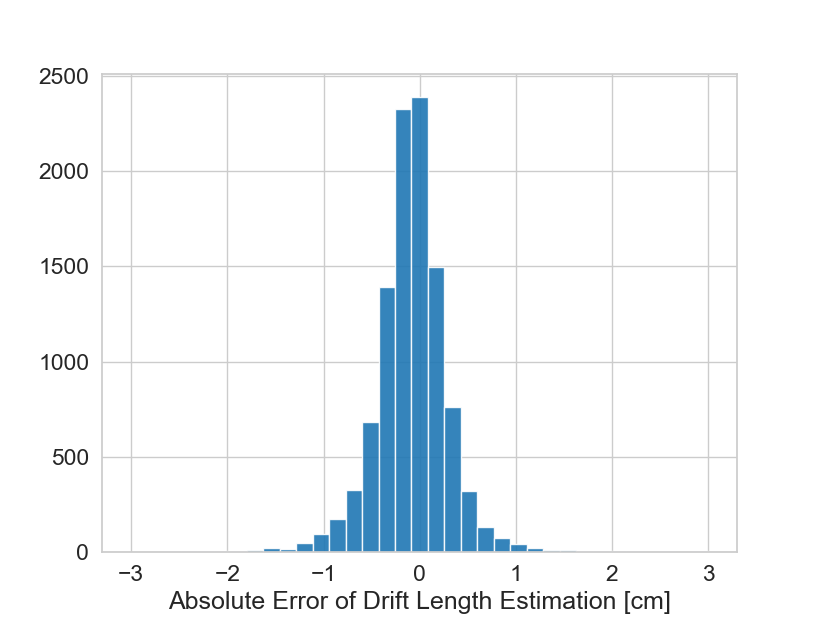}
  \caption{(left) Correlation between true and estimated drift for 500keV electron tracks, and (right) with variance of the measured distance, averaged over all energies and drift distances.}
  \label{fig:drift_diffusion}
\end{figure}

We have since used ML to look at how well the GAMPix technique is able to use the diffusion-independent coarse signals and diffusion-dependent pixel samples to measure the amount of diffusion, and thereby the drift distance. We approached this with a simple neural network with two hidden layers with 64 and 32 hidden neurons, trained on $\sim$ 1000 electron tracks for a range of drift lengths and energies.  In Fig.~\ref{fig:drift_diffusion} we show the there is an excellent correlation between estimated and true drift distances with a modest worsening with distance, and a variance averaged over all energies and depths of $\sim$\,1\,cm. This is a significant capability of the GAMPix architecture. As noted above, the relatively slow electron drift speed in a liquid noble TPC is arguably the primary limitation of the technology, as it limits the rate of particles that can be measured without confusion between events. That confusion arises when a second event occurs before the maximum drift time has passed since the first event, preventing unambiguous matching of the scintillation and charge signals, and thus unambiguous determination of the depth of each track. This independent measure of drift distance resolves the ambiguity.  We expect this to decrease pile-up in GammaTPC by a factor of $\sim\,5$, and we discuss its potential impact in DUNE below.

\section{Possible DUNE application}\label{dune_application}

The Deep Undergound Neutrino Experiment (DUNE) is currently under construction and will be the flagship US on-shore High Energy Physics experiment in the coming decade. DUNE explores neutrino physics at both the GeV and MeV scales.
GeV-scale neutrinos that are produced at the Fermi National Accelerator Laboratory (FNAL) will travel 1300 km to the DUNE Far Detector (FD) in Sanford Underground Research Facility (SURF) in South Dakota. The "long baseline" between FNAL and SURF will enable the definitive measurements of neutrino oscillation. Additionally, MeV-scale neutrinos will be detected from a variety of astrophysical sources, including SuperNova bursts. 

The FD will be composed of four modules, each with a fiducial volume in excess of 10 kTons of LAr. While the design of the first and second modules is defined \cite{dunefd2_2023,dunefd1_2020},
the design of the third and fourth modules has not yet been decided.
While the basic structure of a Liquid Argon Time Projection Chamber (LArTPC) is
assumed, an improved anode readout design that enhances DUNE's physics sensitivity is an attractive possibility. The readout for the second module will be strips on a PCB with length of 2-3 m. Noise is expected to be about 500 e-. \cite{dunefd2_2023}  Through its better noise performance and spatial resolution, the GAMPix architecture could provide significantly better performance. In the following sections, we describe how this performance could impact DUNE and how it could be implemented.


\subsection{Potential impact}
As described in \cite{Friedland_2019} and \cite{Friedland_2020}, a key aspect in improving the performance of the DUNE detector for detection
of GeV-scale neutrinos, is the detection of low energy signals: the so called ``blips'' which are MeV or lower energy electron recoil tracks from the multiple Compton scatters of gamma rays. These are a signature of the various primary decay channels and can represent an important fraction of the energy of the event. They are especially important for neutrons, which deposit a large fraction of their visible energy in the form of gamma rays from the de-excitation of nuclei that have interacted with neutrons, and thus in blips \cite{Friedland_2019}. 
There is also missing energy to nuclear dissociation from neutrons, making it all the more important to measure the neutron energy that is visible.

\begin{figure}[ht]
\centering
\begin{minipage}[b]{0.475\textwidth}
    \includegraphics[width=\textwidth]{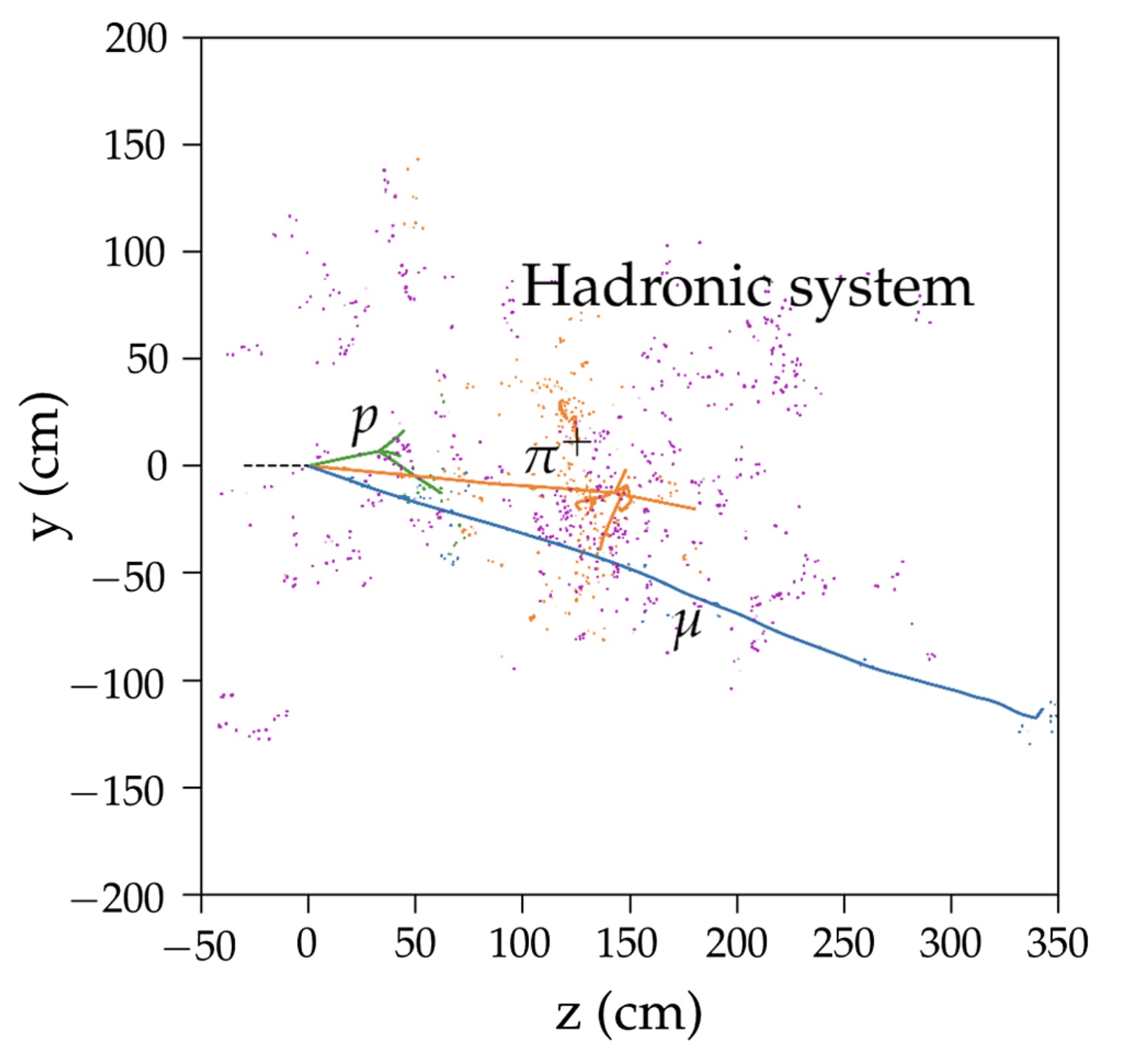}
    \caption{Simulated DUNE GeV-scale neutrino interaction with substantial neutron energy, mostly deposited as gamma rays, which leave a halo of small charge blips, indicated by purple dots. (from \cite{Friedland_2019})}
    \label{fig:DUNE_event}
\end{minipage}
\hfill 
\begin{minipage}[b]{0.475\textwidth}
    \includegraphics[width=\textwidth]{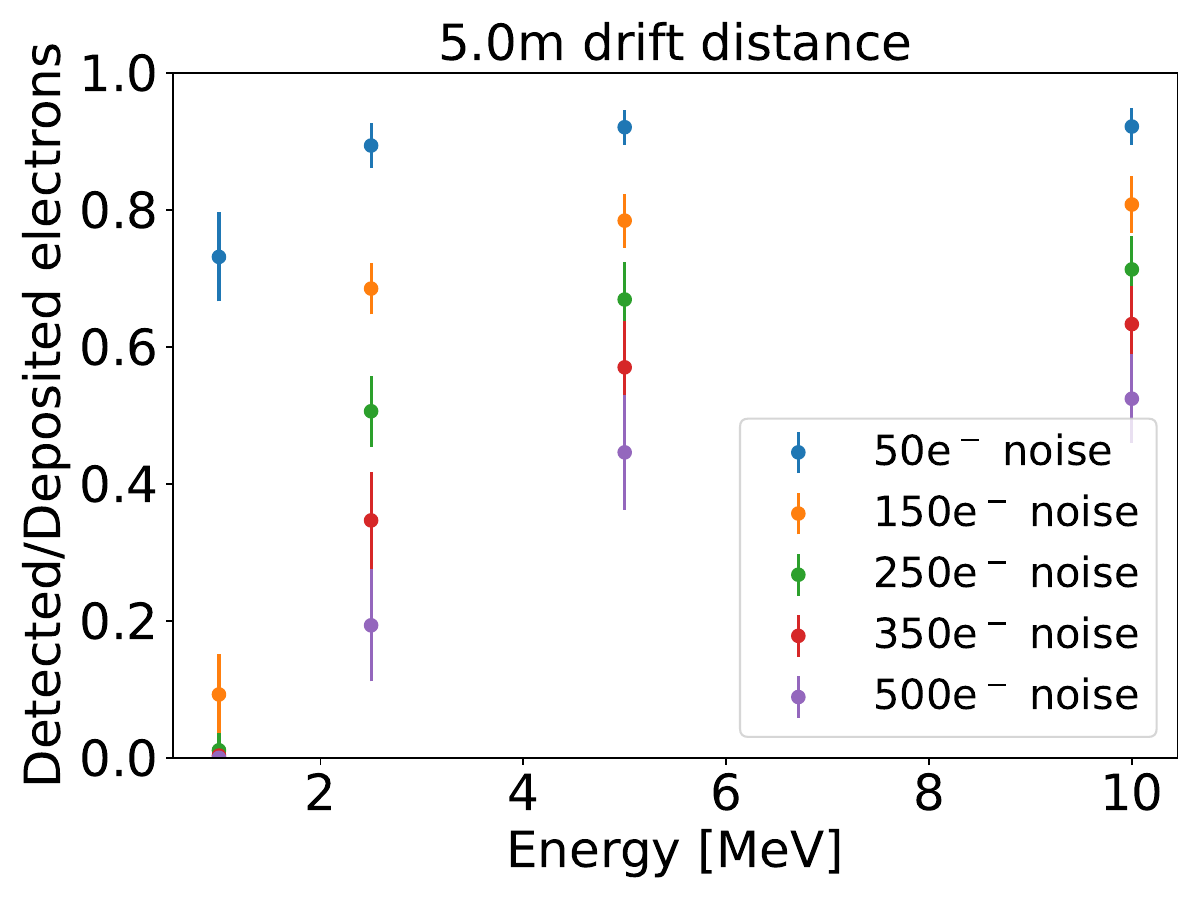}
    \caption{The ratio of electrons detected to those deposited as a function of deposited energy and readout noise. The energy deposition is assumed to occur 5 meters from the anode. As the electrons drift this distance they diffuse and some are collected on pixels that do not collect enough charge to be above threshold. This leads to a serious reduction in detected charge. The pixel size is assumed to be 4mm square, and the electron lifetime 6 ms.}
    \label{fig:DUNE_threshold}
\end{minipage}
\end{figure}

This is seen in Fig.~\ref{fig:DUNE_event}, where the purple blips are the only visible energy originating from neutrons. Note that in this event they dominate the total number of blips, a situation that is typical for several-GeV neutrino events. 

As shown in Fig.~\ref{fig:DUNE_threshold}, the noise level of the readout system greatly impacts the fraction of detected electrons for small energy depositions. This is because the electrons diffuse as they drift, and many of them will be collected on sensors that are below threshold. This means that for the conventional wire-based DUNE charge readout, with noise above 500 e-, these blips are mostly unmeasured. Similarly, if a pixel readout scheme has noise in the 500 e- range, most of the blips will be invisible. In a GAMPix-based LArTPC, with noise on order 50 e-, these blips become visible, and thus enhance energy resolution and overall physics performance.

In addition to improving GeV-scale energy resolution, GAMPix could have multiple benefits for the MeV-scale physics goals of DUNE. As described in Section~\ref{sec:track_imaging}, the GAMPix spatial resolution allows for a measurement of drift length based on diffusion. This would be highly advantageous for DUNE, because the ``start time'' signal, which normally comes from the photon-detection system, may not be present, or be ambiguous, for the small MeV-scale deposits. The indepedent measurement of drift length provides a means for correcting for electron absorption, as well as enabling rejection of background that occurs near the edge of detector (at very short drift time). 

\subsection{DUNE implementation}\label{dune_implementation}

The DUNE FD modules will have drift lengths of approximately 
5m, which means that, due to diffusion, the pixel size should be about 5 mm, rather than the 500 $\mu$m in GammaTPC.  The resolution required for DUNE physics goals is also roughly 5 mm. Furthermore, the noise levels required are somewhat higher than in GammaTPC, due to the larger energy deposits measured in DUNE. In any case, the larger capacitance of the larger pixel size would make acheieving the same noise levels of GammaTPC unfeasible.

Despite these differences, the overall architecture for a DUNE implementation of GAMPix is similar to that of the GammaTPC.   For both applications the readout sensor electrodes are housed in LAr at a temperature of between 87 K for DUNE and up to 120 K for GammaTPC.  The power budget is similar to GammaTPC.  Thus in both applications a charge readout is needed that is simultaneously fine grained, low noise and very low power, with a very large number of sensors housed in a cryogenic fluid.  

\begin{figure}[htbp]
  \centering
  \includegraphics[width=0.8\textwidth]{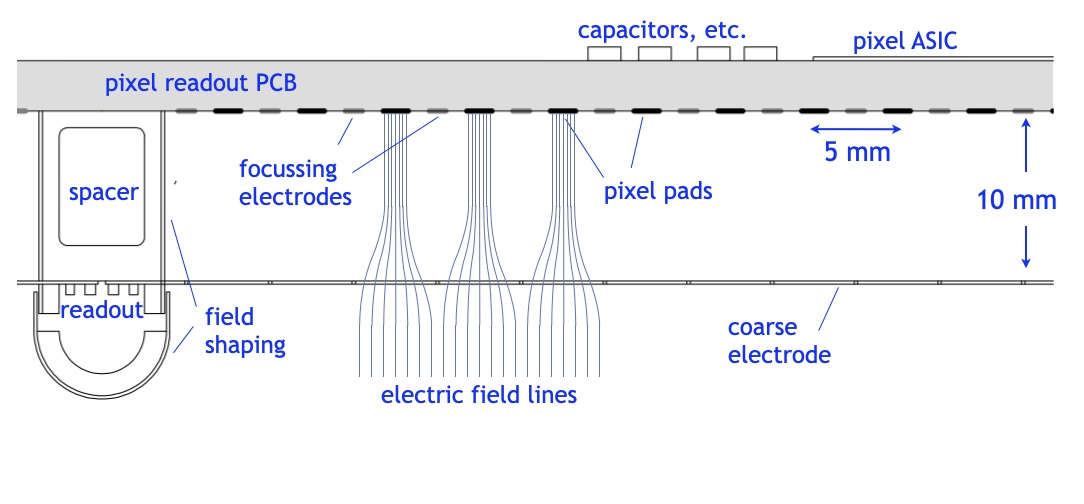}
  \caption{
  \label{fig:DUNE_architecture} A possible architecture for implementation
  of a GAMPix readout in a DUNE FD module. A coarse electrode provides a
  trigger signal for a GAMPix ASIC, which receives signals from 5mm pixels
  on short copper traces.}
\end{figure}

Fig.~\ref{fig:DUNE_architecture} shows a possible detector architecture for implementing the GAMPix scheme in a DUNE FD module. It employs several different techniques as compared to the GammaTPC  design. Since the pixels are larger, it is more convenient to construct them as pads on a printed circuit board (PCB), rather than directly on the ASIC. Short traces would carry the pad signals to a single ASIC per board. The capacitance of the pad and trace system should be about 10pF. Also, instead of wires, an alternative coarse electrode geometry is individually instrumented ''squares'' of wire sized for individual each pixel chip, or small group of pixel chips, with the size chosen as a tradeoff between small size and hence lower capacitance and noise, and higher channel count and thus power. This electrode would be used to trigger the corresponding ASIC, thereby keeping the average power low. 
The main drawback is that each standoff mechanical structure, which must also house the readout electronics, will introduce distortions of the fields from the footprint of a ''readout standoff'' per pixel chip. 
This would be undesirable for the GammaTPC application, although it appears quite feasible in a possible application of GAMPix for DUNE.

This architecture has not yet been studied in as much detail as the
GammaTPC one.  Nonetheless,  we expect the noise to be about 50e-, based on the
studies that have been done for the GammaTPC architecture.
As shown in Figure~\ref{fig:DUNE_threshold}, this low noise level would allow efficient detection of energy deposits below 1 MeV, even at the full drift length of 5 meters. As previously described, this would have a host of benefits for DUNE physics sensitivity. 

\section{Conclusion}

Motivated by the demanding requirements of the GammaTPC gamma ray instrument concept, we have developed a
new charge readout architecture that builds on recent development of large scale, cryogenic ASIC charge readout systems for LAr and LXe TPCs. This innovation features measurements on both coarse and fine scales, which allows diffusion-limited, true 3D pixel imaging while preserving the measurement of the integral charge. It further provides an enormous reduction in power in sparse data environments, with concomitant front-end sparsification of data. Finally, it provides an independent measurement of drift depth, which is invaluable for reducing pile-up in high rate environments.  Simulations of a preliminary design of the crucial power-switching, low-noise front end CSA shows that the requirements for this challenging ASIC can be met.  In addition to enabling a potentially transformative gamma ray detector, this technology has the potential to provide significant improvements to the upcoming modules 3 and 4 of the DUNE Far Detector.  It should also be more broadly applicable to a range of applications where the finest levels of imaging in a TPC are required.

\appendix
\addcontentsline{toc}{section}{Appendices} 
\section{Mathematical Formulation of the Charge Reconstruction Algorithm}\label{appendix:first}

We start with the assumption that the current reconstructed from pixels $I_r(x,y,t)$ is a fraction of the current from the track $I_0(x,y,t)$
\begin{equation}
I_r(x,y,t) = \mu I_0(x,y,t)
\end{equation}
where $\mu < 1$.
Furthermore, due to the linearity of the response of the wire grid (at position i,j) $R^{ij}(x,y,t)$, (defined as Electric Field Response Function for a wire i,j) we can get the signal on every wire $S_{0}^{ij}(t)$ by convolving of the response function with the current along the time axis. The 0 index denotes the signal from the original track (either in hardware or simulation).

\begin{equation}
S_{0}^{ij}(t) = R^{ij}(x,y,t) \otimes_t I_0(x,y,t)
\end{equation}

Similarly, we can produce the wire signals $S_{r}^{ij}(t)$ from the reconstructed current $I_r(x,y,t)$,

\begin{align}
S_{r}^{ij}(t) &= R^{ij}(x,y,t) \otimes_t I_r(x,y,t) \\
&=R^{ij}(x,y,t) \otimes_t \mu I_0(x,y,t)\\
&= \mu S_{0}^{ij}(t)
\end{align}

From there, we can express the ratio of the currents $I_r(x,y,t)$ and $I_0(x,y,t)$ as the ratio of the signals on the wires:
\begin{equation}
\mu = \frac{S_{r}^{ij}(t)}{S_{0}^{ij}(t)}
\end{equation}

With sampled signal that contains noise, we will get the best estimate of the ration by using the result of the least squares method:

\begin{equation}
\mu = \frac{\sum_{aw}\sum_{k} S_{r}^{ij}[k] S_{0}^{ij}[k] }{\sum_{aw} \sum_{k} S_{0}^{ij}[k]S_{0}^{ij}[k]} 
\end{equation}

The sum $\sum_{aw}$ runs over all activated wires while $\sum_{k}$ runs over signal that is above the threshold. That information is extracted from the simulated wire readout $S_{r}^{ij}(t)$ by applying the trigger (described above) to it. It is more practical to take this ration in the fourier domain together with a wiener filter to suppress higher frequencies. 

To get the total charge of the track, we integrate the current over space and time:
\begin{align}
q_o &= \iiint I_0{(x,y,t)}v_d\,dx\,dy\,dt \\
&= \frac{1}{\mu} \iiint I_r{(x,y,t)}v_d\,dx\,dy\,dt \\
&= \frac{q_r}{\mu}
\end{align}
yielding that the ratio of currents is equal to the ratio of charges. Since we have $q_r$ from the pixel readout, and the quotient $\mu$ from the wire signals, we can recover the total charge of the track.


\acknowledgments

This work was supported by NASA grant NNH21ZDA001N-APRA.  We thank Gunther Haller for useful discussions on the pixel ASIC architecture.


\bibliographystyle{ieeetr} 

\bibliography{main.bib}

\end{document}